\documentclass[pr1,amsmath,twocolumn,showkeys,superscriptaddress,preprintnumbers]{revtex4}
\usepackage[utf8]{inputenc}
\usepackage{graphicx}
\usepackage{setspace}
\usepackage{amsmath}
\usepackage{amssymb}
\usepackage{dsfont}
\usepackage{color}
\usepackage{url}
\allowdisplaybreaks
\usepackage[nobiblatex]{xurl}

\newcommand {\nc} {\newcommand}
\nc {\cgmf} {$\texttt{CGMF}$}
\nc {\gray} {$\gamma$ ray}
\nc {\grayd} {$\gamma$-ray}
\nc {\nup} {$\bar{\nu}$}
\nc {\nug} {$\bar{N}_\gamma$}
\nc {\pnu} {$P(\nu)$}
\nc {\pnug} {$P(N_\gamma)$}

\begin{document}
\title{A novel emulator for fission event generators}
\author{ Karl Daningburg }
\email{kd1956@rit.edu}
\affiliation{ School of Mathematics and Statistics, Rochester Institute of Technology, Rochester, New York, 14623}
\affiliation{ Theoretical Division, Los Alamos National Laboratory, Los Alamos, New Mexico, 87545 }
\author{ A.~E.~Lovell }
\email{lovell@lanl.gov}
\affiliation{ Theoretical Division, Los Alamos National Laboratory, Los Alamos, New Mexico, 87545 }
\author{ R.~O'Shaughnessy}
\affiliation{ School of Mathematics and Statistics, Rochester Institute of Technology, Rochester, New York, 14623}

\date{\today}
\preprint{LA-UR-24-24664}

\begin{abstract}
    A wide variety of emulators have been developed for nuclear physics, particularly for use in 
    quantifying and propagating parametric uncertainties to observables.  Most of these methods have been used 
    to emulate structure observables, such as energies, masses, or separation energies, or reaction observables, 
    such as cross sections.  Rarely, if ever, have event generators for theory models been emulated.  Here, we 
    describe one such novel emulator for the fission fragment decay code, \cgmf{}, which calculates the emission 
    of prompt neutrons and \gray{}s from fission fragments.  The emulator described in this work uses a combination of a noisy emission
    model and mixture density network to model the neutron and \grayd{} multiplicities and energies.  In this 
    manuscript, we display the power of this type of emulator for not only modeling average prompt fission 
    observables but also correlations between fission fragment initial conditions and these observables, using 
    both neutron-induced and spontaneous fission reactions.
\end{abstract}


\maketitle

\section{Introduction}
\label{sec:intro}

In a fission event, a heavy nucleus splits into two---or more---lighter fission fragments that are highly 
excited and very quickly decay through neutron and \grayd{} emission, so-called prompt emission.  On longer time scales, 
these resulting nuclei further decay toward stability, so-called delayed emission.  Average quantities from these 
decays, such as the multiplicities and energies of the neutrons and \gray{}s, along with spectra, are used 
to understand fission properties and in applications, e.g. \cite{Jaffke2018,Randrup2019,Talou2018}; however, correlated fission 
observables can give further insight to the underlying physics and are also needed for performing 
detailed simulations, such as those in transport codes.  When studying fission theoretically, 
the calculations are often broken down into three stages.  The first describes the initial deformation and 
splitting of the fissile nucleus, typically through microscopic or microscopic-macroscopic models \cite{Schunck2014,Sierk2017,Verriere2019,Regnier2016,Moller2015,Usang2019,Ishizuka2017,Mumpower2020,Verriere2021}.  
These methods have been used to determine fission fragment masses, charges, and kinetic energies, 
as well as to gain insight into the spins of the fission fragments \cite{Randrup:2014,Marevic2021,Bulgac2021,Randrup:2021,Stetcu2021}.  After scission, 
separate models are used to de-excite the fission fragments through prompt neutron and \grayd{} emission, 
\cite{CGMF,FREYA1,FREYA2,FIFRELIN,Okumura2018,Lovell2021,GEF,Tudora2017}.  Finally, the 
delayed emission calculations can be solved as time-dependent or time-independent manors, often using available evaluated decay data, such as from the ENDF/B \cite{ENDFB8}, 
JEFF \cite{JEFF33}, or JENDL \cite{JENDL5} nuclear data libraries.  To date, there is no complete theory of nuclear fission.

The Hauser-Feshbach fission fragment decay code \cgmf{}, developed at Los Alamos National Laboratory, models 
the second stage of the fission process as described above \cite{CGMF}: the emission of the prompt 
neutrons and \gray{}s.  Fission fragment initial conditions in mass, charge, total kinetic energy, spin, 
and parity, $Y(A,Z,\mathrm{TKE},J,\pi)$, are phenomenologically parametrized and fit to available 
experimental data.  For more details on the exact parametrization, see \cite{CGMF}.  For each event, these 
five quantities are sampled, then the total excitation energy (TXE) is calculated for the given fragmentation 
based on conservation of energy, $\mathrm{TXE} = \mathrm{TKE} - Q$, where $Q$ is the Q value for the fragment 
split.  The TXE is shared between the fragments based on a ratio of temperatures, $R_T$, and the TKE is 
shared based on momentum conservation.  Once the initial conditions are determined for each fragment, the 
Hauser-Feshbach statistical theory \cite{HauserFeshbach} is used to de-excited each fission fragment.  
Neutrons and \gray{}s are emitted probabilistically, based on the combination of the level density and 
transmission coefficient calculated from the Hauser-Feshbach theory.  Numbers of neutrons and \gray{}s, their energies and directions, and the initial conditions of the fission fragments are 
collected and can be post-processed to study mean observables and correlations.

To begin the Hauser-Feshbach calculation, each fission fragment is described by its mass, $A$, charge, $Z$, kinetic energy, 
$\mathrm{KE}$, excitation energy, $U$, spin, $J$, and parity, $\pi$.  At each step in the decay, energy, momentum, spin, 
and parity are conserved. Although these six values are enough to describe the initial conditions of the fission fragments, 
the full decay---including the probabilistic nature of \cgmf{}---is much more complicated, depending on how these initial conditions 
change as the neutrons and \gray{}s are emitted.  We also note that \cgmf{} can run in two modes.  In the first, the full 
Hauser-Feshbach decay is calculated.  In the second, only the fission fragment initial conditions are sampled, without performing 
the neutron and \grayd{} emission.  This second mode runs significantly faster than the full Hauser-Feshbach decay; this speed is useful to connect with the emulator 
we developed, as will be described in this manuscript.

Although \cgmf{} is very powerful for studying correlated fission observables, it must be run on High 
Performance Computing resources when large analyses are to be completed.  Due to these computational challenges across nuclear physics, emulators have become increasingly 
common, particularly as the number of studies on 
parametric uncertainty quantification has grown.  Gaussian process (GP) emulators \cite{Novak2014,Sangaline2016,Melendez2019,Surer2022} have 
recently been replaced with reduced basis methods (RBMs) \cite{Frame2018,Sarkar2021,Koenig2020,Ekstrom2019,Furnstahl2020,Wesolowski2021}.  
Unlike the GPs, where the emulator is a non-linear mapping from some set of inputs to one (or more) outputs, 
the RBMs use a small set of basis functions to emulate the Schr\"{o}dinger equation directly.  Because the 
Schr\"{o}dinger equation itself is emulated, instead of the resulting observables such as energies or cross 
sections, the extrapolations from RBMs tend to be much more robust than those from GPs, as in
\cite{Koenig2020}.  
Neural networks have also been used for performing uncertainty 
quantification for theory models \cite{Utama2016,Utama2016a,Neufcourt2018,Wang2019,Neufcourt2020a}.  
However, in nuclear physics, emulators rarely, if ever, have been used for event generators.  

In this work, we aim to create an emulator for \cgmf{} that significantly speeds up the Hauser-Feshbach 
piece of the calculation while retaining correlations between fission fragment initial conditions and 
prompt fission observables.  Part of the calculation of the neutron energies and the competition between neutrons 
and \gray{}s involves solving the Schr\"{o}dinger equation for the neutron transmission coefficients, so RBMs 
could be used to speed up this piece of the calculation.  We note that these methods have been used for 
cross section calculations \cite{Drischler2021,Odell2024}; however, just implementing this method would not speed up 
the initial condition sampling or any calculations for the \grayd{} properties.  Additionally, because we 
want our emulator to retain the correlations between prompt neutron and \grayd{} observables and the fission 
fragment initial conditions, we cannot just emulate average multiplicities or energies.  Therefore, we 
develop a novel emulator that combines stochastic and probabilistic machine learning techniques that provides event-by-event emulation.

This manuscript is divided into the following sections.  In Sec. \ref{sec:methods}, we describe our stochastic 
and probabilistic methods used for the emulation of prompt neutron and \grayd{} multiplicities and energies.  
In Secs. \ref{sec:results:mult} and \ref{sec:resultsEnergy}, we show the results from the multiplicity and energy emulators.
We discuss potential uses of the full emulator in Sec. \ref{sec:fullEmulator}.
 Finally, in Sec. 
\ref{sec:conclusion}, we summarize our results and describe the use of these types of emulators and how 
they can be improved.

\section{Methods}
\label{sec:methods}

Our emulators for the neutrons and \gray{}s that are emitted during the decay of a single fission fragment involve two stages.  In the first stage, we predict the multiplicity, $\nu$ and $N_\gamma$, of neutrons and
\gray{}s that are emitted.  The structure of these emulators is described in general in \ref{sec:multiplicity}, while section \ref{sec:designChoices}
provides the concrete training data and dimensional reduction choices  we adopt to realize this emulator.
In the next stage, we  use a
conditional probabilistic emulator, described in Sec. \ref{sec:MDN}, to predict the neutron and \grayd{} energies for each one
emitted, $E_n$ and $E_\gamma$, given the assumed-known multiplicities of both types of particles.  

\subsection{Multiplicity Emulator}
\label{sec:multiplicity}

\subsubsection{Stochastic Emulator Design} 
\label{sec:design}
In fission, the outcomes for multiplicity are effectively a multinomial distribution, a 
generalization of the binomial distribution to $n$ outcomes. Concretely, a single 
set of inputs may lead to multiple discrete outcomes. The data reflects this 
randomness; however, since the inputs to our model are also generated according 
to the stochastic nature of fission, we do not have multiple datapoints for each 
unique set of inputs. In fact, since some inputs, such as the excitation energy, $U$, are continuous, 
it is a virtual certainty that each datapoint is unique. Practically speaking, 
this means that we do not have direct access to the multinomial distribution at any 
point of the input space. The construction of our emulator must capture the stochasticity 
inherent in the problem, while being trainable on single multiplicity values at each datapoint.

We describe a noisy emission model (NEM) for the neutron and \grayd{} multiplicities.  To simplify our discussion, in this methods section we assume a
maximum multiplicity of $\nu=4$, suitable for neutrons, but we employ the same algorithm with a larger range of
multiplicities for \gray{}s.   We begin by constructing 
a probability density function for a continuous parameter, $n$, ranging from 0 to 4, which covers the 
range for multiplicity outcomes we are emulating: 0, 1, 2, 3, or 4 neutrons emitted. Assuming a Gaussian 
form for this parameter, we must learn the mean, $\mu$, and standard deviation, $\sigma$, of $n$ as a function 
of input vector, $\mathbf{x}$:
\begin{equation}
    f_n(n|\mathbf{x}) = \mathcal{N} (\mu(\mathbf{x}), \sigma(\mathbf{x}))
\end{equation}
From $f_n(n|\mathbf{x})$ we estimate the probability of realizing discrete value $\nu$ from 
$\mathbf{x}$ using the standard normal cumulative distribution function $\Phi$:
\begin{align}
    \label{eqn:p0}
    \texttt{P}_\nu(0|\mathbf{x}) &= \Phi\left( \frac{0.5 - \mu(\mathbf{x})}{\sigma(\mathbf{x})}\right),\\
    \texttt{P}_\nu(1|\mathbf{x}) &= \Phi\left( \frac{1.5 - \mu(\mathbf{x})}{\sigma(\mathbf{x})}\right) - \Phi\left( \frac{0.5 - \mu(\mathbf{x})}{\sigma(\mathbf{x})}\right),\\
    \texttt{P}_\nu(2|\mathbf{x}) &= \Phi\left( \frac{2.5 - \mu(\mathbf{x})}{\sigma(\mathbf{x})}\right) - \Phi\left( \frac{1.5 - \mu(\mathbf{x})}{\sigma(\mathbf{x})}\right),\\
    \texttt{P}_\nu(3|\mathbf{x}) &= \Phi\left( \frac{3.5 - \mu(\mathbf{x})}{\sigma(\mathbf{x})}\right) - \Phi\left( \frac{2.5 - \mu(\mathbf{x})}{\sigma(\mathbf{x})}\right),\\
    \label{eqn:p4x}
    \texttt{P}_\nu(4|\mathbf{x}) &= 1 - \Phi\left( \frac{3.5 - \mu(\mathbf{x})}{\sigma(\mathbf{x})}\right).
\end{align}
We wish to maximize the probability of realizing the correct value $\nu_k$ for each $\mathbf{x}_k$, 
where the correct mapping $\mathbf{x}_k$ to $\nu_k$ is determined by the training set constructed from the full \cgmf{} calculations. 
Here $k$ represents a single point in the training set.

To maximize this probability, we let 
$f_n(n | \mathbf{x})$ be a neural network with hyperparameters $\phi$ and outputs $\sigma(\mathbf{x})$ and $\mu(\mathbf{x})$: $f_n(n | \mathbf{x}) = f(\mathbf{x};\phi)$.
We define the log likelihood, $\ln \mathcal{L}$,
\begin{align}
    \ln \mathcal{L} &=   \sum_{\nu_k=0} \ln \texttt{P}(0|\mathbf{x}_k) +   \sum_{\nu_k=1} \ln \texttt{P}(1|\mathbf{x}_k) \nonumber \\ &+\sum_{\nu_k=2}
    \ln \texttt{P}(2|\mathbf{x}_k)  +\sum_{\nu_k=3}\ln \texttt{P}(3|\mathbf{x}_k) \nonumber \\ &+ 
    \sum_{\nu_k=4}\ln \texttt{P}(4|\mathbf{x}_k),
\end{align}
where probabilities $\texttt{P}(n|\mathbf{x})$ are given by Eqs. (\ref{eqn:p0})--(\ref{eqn:p4x}). 
We define the loss function
\begin{equation}
    \mathbf{L} (\mathbf{x},n, \phi) = - \ln \mathcal{L} .
\end{equation}
We learn the parameters $\phi$ by minimizing this loss function. This is equivalent to 
maximizing the log likelihood of realizing the neutron multiplicities in the dataset $\nu_k$ at each 
input $\mathbf{x}_k$ with $f(\mathbf{x}_k;\phi)$.

\subsubsection{Multiplicity training data}
\label{sec:multTraining}

We focus our emulator on spontaneous fission and first-chance fission for neutron-induced reactions (that 
is, below about 5 MeV incident neutron energy where a neutron has a near zero probability of being emitted from the compound nucleus before 
it fissions).  In our training set, we include events from $^{252}$Cf spontaneous fission as well as 
$^{235}$U(n,f), $^{238}$U(n,f), and $^{239}$Pu(n,f) at incident neutron energies of thermal, 1 MeV, 2 MeV, 
3 MeV, 4 MeV, and 5 MeV. We train on 100,000 events from each of these 19 fission reactions.  Along with using our emulator to predict events from these reactions, we additionally predict 
other fission reactions available within \cgmf{}, including $^{233}$U(n,f), $^{234}$U(n,f), $^{237}$Np(n,f), 
$^{241}$Pu(n,f), again up to 5 MeV incident neutron energy.

\subsubsection{Multiplicity emulator design choices}
\label{sec:designChoices}

The NEMs describe in Sec. \ref{sec:design} provide a surrogate model, $\hat{\nu}$, for neutron (or similarly \grayd{}) multiplicity. 
While six physical quantities that define the initial conditions of the fission fragments, $A$, $Z$, KE, $U$, $J$, and $\pi$, are available as inputs for our model, 
the prompt observables do not have a strong sensitivity to all of these initial conditions.  
For example, when modeling prompt decay products, the $U$ available 
 in a fission fragment determines the number of neutrons and \gray{}s that can be emitted. 
Because the kinetic energy of the fragments 
 are highly correlated to the excitation energy, 
 including KE and $U$ into the input vector 
 is likely repeated information. We confirmed that including kinetic energy as an input is detrimental to performance 
 by training and testing such a network.

In \cgmf{}, the initial parity of each fission fragment is sampled randomly, split evenly between positive and negative.  
For many of the \grayd{} emissions, the $J$ and $\pi$ of the initial and surrounding nuclear states determines whether a decay is 
accessible or not.
Parity, however, plays a much smaller role in the neutron emission, as the underlying equations average over the two parity states.  
Therefore, in our emulators we use $\pi$ as an input for the \grayd{} multiplicity emulator but not for the neutron emulator.

In our two emulators, we therefore consider $\textbf{x} = \{A,Z,U,J\}$ for the neutron multiplicity emulator 
and $\textbf{x} = \{A,Z,U,J,\pi \}$ for the \grayd{} multiplicity emulator.
Having selected physically motivated fission fragment initial conditions from \cgmf{} 
for the two emulators, we now design a model 
$\hat\nu$ capable of representing the stochastic nature of the decay product multiplicity.

To construct $\hat{\nu}$, we tested fully connected neural networks with ReLU activation functions
of varying architectures: 2, 4, and 8 hidden layers with 16, 32, and 64 nodes each. For neutrons the highest performing architecture
was the 4 layer 64 node model, while for \gray{}s an 8 layer 64 node model performed best.
Inputs were normalized to have means of 0 and standard deviations of 1.
Two output nodes represented a Gaussian's mean $\mu$ and standard deviation $\sigma$ for parameter $n$, 
as described in Sec. \ref{sec:design}.

\subsubsection{Validation Strategies} \label{sec:validate}
Having outlined a strategy for training a multiplicity emulator on the raw data available to us, 
we now require methods for validating the stochastic emulator $\hat{\nu}$ versus $\nu$. 
These validation tests require comparing some empirical distribution of $\hat{\nu}$ to some empirical distribution of
$\nu$ using a sufficiently large sample of training data.  In App. \ref{ap:event}, we describe concrete methods to select
suitable narrow subsamples and to comprehensively quantify differences between these empirical distributions.  For
simplicity, in the main text, we will assume a suitably large and consistent number of subsamples have been identified. When quantifying the difference between NEM and \cgmf{}, we
will primarily employ simple summary statistics (the sample mean) to compare distributions.

Specifically, we will to compare two distributions' 
\textit{expected values} $\bar \nu$, the average prompt neutron multiplicities, and compute the difference between them.
To get an error metric from expected value, we simply compute 
expected values for both the reference distribution from \cgmf{} $\bar \nu$ and the NEM distribution $\bar{\hat{\nu}}$ and find the difference:
\begin{equation}
    \label{eqn:delta}
    \Delta \bar \nu = \bar \nu - \bar{\hat{\nu}}
\end{equation}
We then report the error as $\Delta \bar \nu / \bar \nu$, in percent. The resulting error metric is a physically meaningful value: the percent error 
in expected average neutron multiplicity of the NEM, $\bar{\hat{\nu}}$, from the full \cgmf{} calculation, $\bar \nu$. 

\subsection{Energy Emulator}
\label{sec:MDN}

\subsubsection{Mixture Density Networks}

The Mixture Density Network (MDN) is a probabilistic machine learning method~\cite{MDNs}.  We use a MDN to 
learn the neutron and \grayd{} energies of each \cgmf{} event, instead of using a standard feed-forward 
neutral network (NN).
Standard deterministic NNs have 
difficulty predicting a distribution of output values from the same input value (as discussed in Sec. \ref{sec:design}); this is 
precisely the case we have for prompt emission from fission fragments, where the same fission fragment initial 
conditions can give rise to multiple neutrons and/or \gray{}s each with a distinct energy.  

The MDN describes a model
output as a mixture of Gaussian functions
\begin{equation}
\textbf{y}(\textbf{x}) = \sum \limits _{i=1} ^{m} \alpha_i (\textbf{x}) \mathcal{N} [ \mu_i (\textbf{x}), \sigma_i (\textbf{x})],
\end{equation}
where $\mathcal{N}$ is the normal distribution whose weights, means, and standard deviations, $\alpha_i 
(\textbf{x})$, $\mu_i (\textbf{x})$, and $\sigma_i (\textbf{x})$, are learned by a feed-forward NN.  The 
total number of Gaussian mixtures is $m$.  This mixture of Gaussians allows for the full posterior 
distribution of the model to be described, instead of having to assume a fixed shape for a given set of input parameters.

The loss function is defined as
\begin{equation}
\mathcal{L} = - \ln {\left [ \sum \limits _{i=1} ^m \frac{\alpha_i(\textbf{x})}{(2\pi)^{1/2} \sigma_i (\textbf{x})} \mathrm{exp} \left \{ -\frac{|| \textbf{t}-\mu_i(\textbf{x}) ||^2}{2\sigma_i (\textbf{x})^2} \right \} \right ]},
\label{eqn:logloss}
\end{equation} 
where $\textbf{t}$ is the vector of training outputs.  This loss function minimizes the difference in 
distributions between the training data and the predicted posterior distribution.  See \cite{Lovell2020} for more detail.

Similar to previous uses of MDNs by one of the authors~\cite{Lovell2020,Lovell2022}, our MDN is written 
in $\texttt{pytorch}$~\cite{pytorch} and can be run on both CPU and GPU, with a significant speed-up found 
on GPU.  

\subsubsection{Energy training data and network architecture}
\label{sec:MDNtraining}

For the two MDNs, we construct our training set using the same 19 reactions as listed in 
Sec. \ref{sec:multTraining}.  \cgmf{} histories of 1 million events for each reaction were 
calculated, and each neutron or \grayd{} energy from a fission 
fragment is considered a single training point.  For both MDNs, 80\% of the energies were 
included in the training set.  All events were used to validate the results.  

For the neutron MDN, each input and output vector was scaled to have a mean of 0 and 
a standard deviation of 1.  That is, we subtract the mean of the training set and divide by 
the standard deviation.  The scaling for the \gray{}s was performed slightly differently.  
Because of the much wider distribution of \grayd{} multiplicities, their skew toward 
higher multiplicities, and the flatter distribution of \grayd{} energies (compared to neutron energies),
 we scale the inputs and output for the \grayd{} MDN to lie between 0 and 1.
Here, we subtract the minimum value from the training set then 
divide by the difference between the maximum and minimum value.

The input of the training set for the neutron MDN included $\textbf{x} = \{ A, Z, \mathrm{KE}, U, J \}$, 
and the input for the \grayd{} MDN included all six fission fragment initial conditions, 
$\textbf{x} = \{ A, Z, \mathrm{KE}, U, J, \pi \}$.  We include KE in the the MDN input because we are training 
on the energies of the neutrons and \gray{}s in the laboratory frame.  Although energy 
available for the decay is given by the excitation energy, the kinetic energy is necessary 
to know the conversion from the center of mass frame to the laboratory frame.  Additionally, 
as discussed in Sec. \ref{sec:designChoices}, we include the parity as an input for the \grayd{} 
emulators due to its importance in the selection of the decay mode.  

For the neutron MDN, our network was fully connected, consisting of 1 layer with 24 nodes 
each and 6 Gaussian mixtures.  The MDN for the \grayd{} energy emulator consisted of 
8 layers with 48 nodes each and 8 Gaussian mixtures.   We trained for 1000 (2500) epochs for the neutron (\grayd{}) 
emulator.  For both MDNs, we tested a variety of Gaussian mixtures, layers, and nodes per layer.  
We did not find a significantly improved performance with larger networks for either emulator.  

\section{Results for the Multiplicity Emulation}
\label{sec:results:mult}

\subsection{Neutron multiplicity}
\subsubsection{Event-by-Event $\Delta \bar \nu$}
\label{sec:event}
\begin{figure*}
    \includegraphics[width=0.9\textwidth]{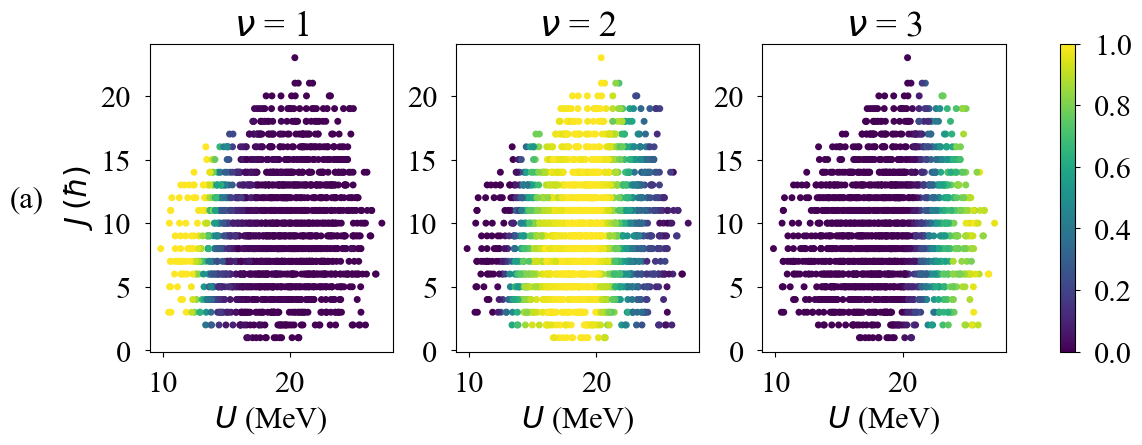}
    \includegraphics[width=0.9\textwidth]{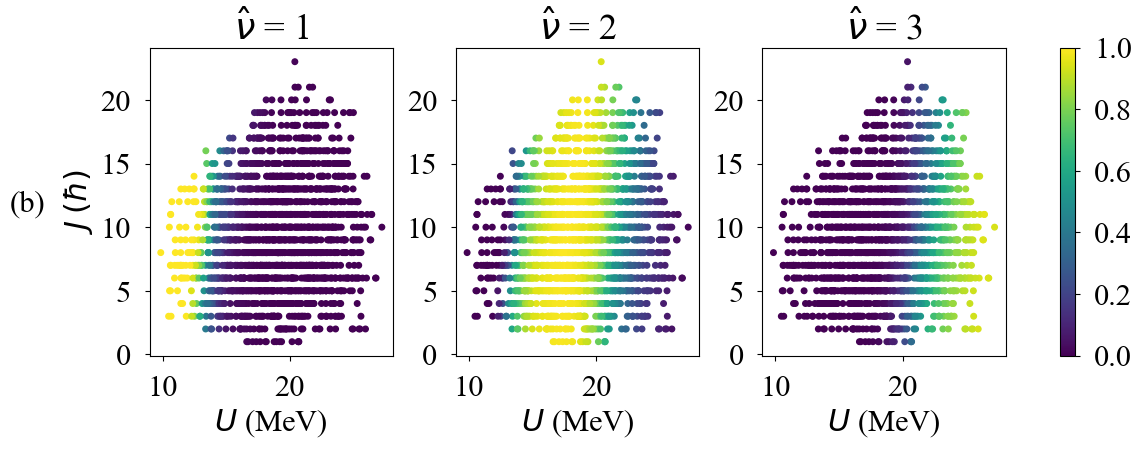}
    \includegraphics[width=0.9\textwidth]{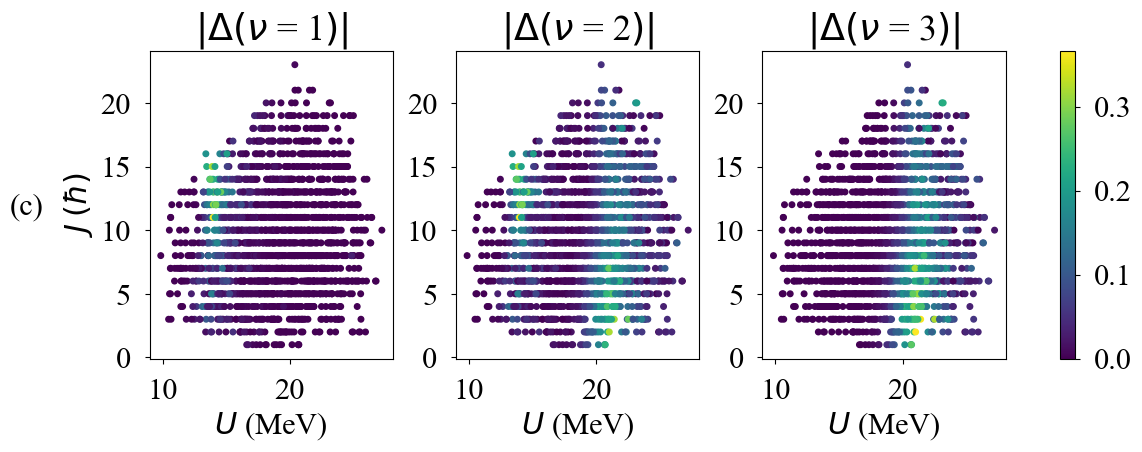}
    \caption{Validating our model for $^{106}$Mo across the possible values for excitation energy, $U$, and spin, $J$. 
    Here $^{106}$Mo is the compound nucleus before neutron and \grayd{} emission. Panel (a) shows the probability of realising
    multiplicities $\nu$ for individual fission events in the $U$-$J$ domain. Panel (b) shows the performance of the emulator $\hat\nu$ 
    at these validation points, and panel (c) shows the difference $\nu-\hat\nu$.}
    \label{fig:validate}
\end{figure*}

For the most resolved and conservative estimate of our emulator's performance, we begin by comparing the emulator to
\cgmf{} on a fragment-by-fragment basis.
To estimate the fragment-by-fragment performance of our model, we use the validation dataset constructed according
to App. \ref{ap:event}.
In Fig. \ref{fig:validate}, we compare the probabilities of
1, 2, and 3 neutrons being emitted from $^{106}$Mo as a function of fission fragment $U$ and $J$ between the neutron NEM and \cgmf{}. As $U$ increases, the peak 
in the probability distribution moves toward higher multiplicities (first row of Fig. \ref{fig:validate}); this is as expected since there is more energy available above the neutron separation energy for more neutrons to be emitted. The same trend is observed in our emulator (second row of Fig. \ref{fig:validate}). The third row of Fig. \ref{fig:validate} shows the absolute difference between the distributions from \cgmf{} and the neutron NEM.  

We note that Fig. \ref{fig:validate} only shows the performance of the emulator for one of many fission fragments in the 
dataset. Next we will introduce a method to summarize the performance on all fission fragments in a given dataset in Sec. 
\ref{sec:global}.

\subsubsection{Global Performance}
\label{sec:global}
We examine global performance by evaluating our emulator on a per-fission-reaction basis, e.g. a single target 
nucleus at a single incident neutron energy (or for spontaneous fission).
It is important to note that global performance will always be better than fragment-by-fragment performance due to the averaging out of local errors.

\begin{figure}
    \includegraphics[width=0.45\textwidth]{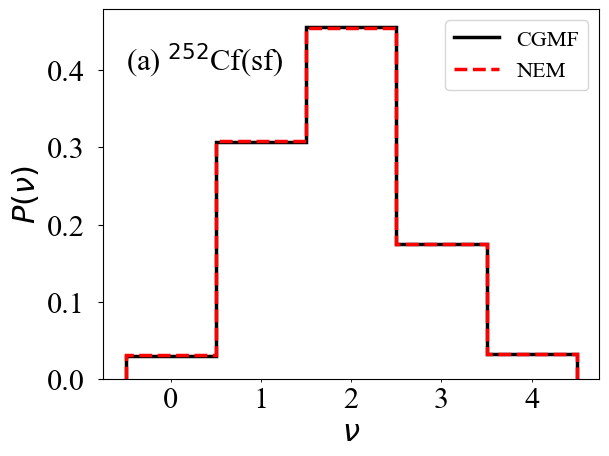}
    \includegraphics[width=0.45\textwidth]{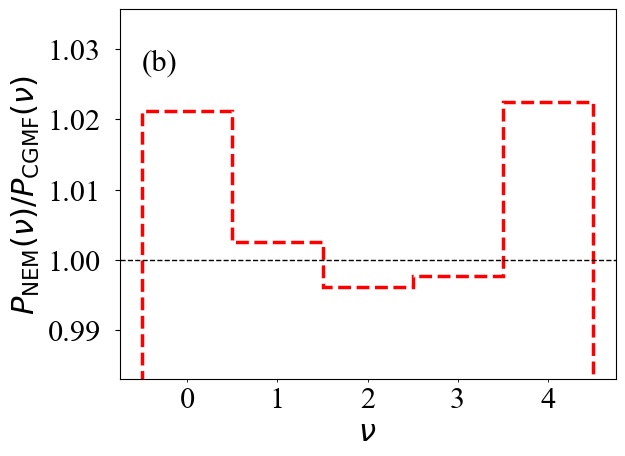}
    \caption{(a) Comparison of the neutron multiplicity distribution for $^{252}$Cf between \cgmf{} (solid black) and the emulator (dashed red). 
  (b) Ratio of the prediction to the empirical value.
}
    \label{fig:cf252_dist}
\end{figure}

To test global performance, we use a test dataset larger than that used for training. The test set was comprised of 1 million fragments per reaction,
whereas the training set was a subset containing 20\% of the test set.  
We separate our test dataset according to the target nucleus and 
 also by the incident neutron energy for the neutron-induced fission reactions. 
 We compare the prompt neutron multiplicity distributions, $P(\nu)$, between the NEM and \cgmf{} for each reaction.  
Figure \ref{fig:cf252_dist}(a) shows $P(\nu)$ for the NEM (red dashed) and \cgmf{} (black solid) for the spontaneous fission
of $^{252}$Cf. 

\begin{figure*}
    \includegraphics[width=0.45\textwidth]{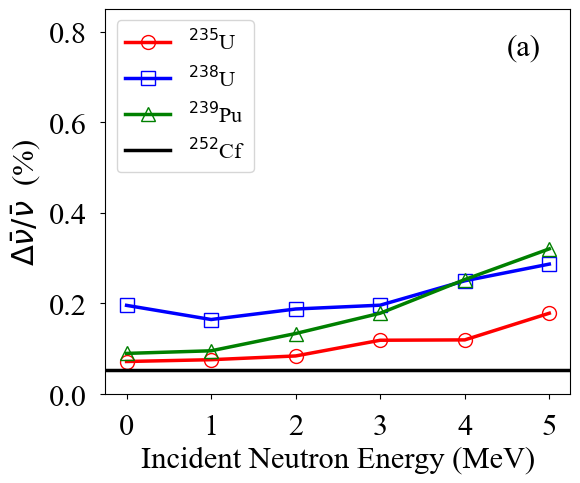}
    \includegraphics[width=0.45\textwidth]{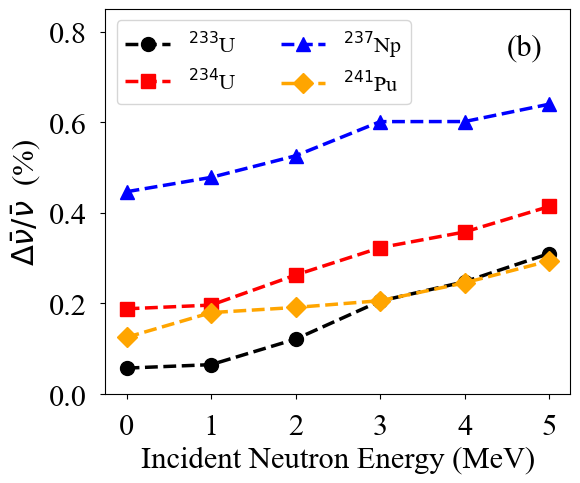}
    \caption{(a) Percent error in the average neutron multiplicity, $\bar{\nu}$, between \cgmf{} and the NEM emulator,
    as in Eq. (\ref{eqn:delta}), for $^{235}$U(n,f) (red), $^{238}$U(n,f) (blue), and $^{239}$Pu(n,f) (green) as a function of incident neutron energy. Error for spontaneous fission 
    of $^{252}$Cf(n,f) is shown in black.
    (b) Same as (a) for reactions that were not included in the training set:  $^{233}$U(n,f) (black), $^{234}$U(n,f) (red), $^{237}$Np(n,f) (blue), and $^{241}$Pu(n,f) (orange).}
    \label{fig:delta_nubar_multi}
\end{figure*}

To better characterize the small differences between our emulator and full \cgmf{} calculation, Fig. \ref{fig:cf252_dist}(b) shows the ratio
between $P(\nu)$ of the NEM and \cgmf{}.
In this case, our emulator is within about 2\% of \cgmf{} for every value of $\nu$, although the differences are better than 0.5\% in the peak of the $P(\nu)$ distribution. 
The largest differences between the emulator and \cgmf{} are seen for $\nu=0$ and $\nu=4$, multiplicities which
only occur in about 3\% of $^{252}$Cf(sf) fission events. Figure \ref{fig:delta_nubar_multi}(a) summarizes the global performance 
of the emulator for a greater number of nuclei, illustrating
$\Delta \bar \nu / \bar \nu$ for each of the 18 neutron-induced fission reactions trained on. Error falls below 0.4\% for all fission 
reactions within the training set. 

To show the generality of the emulator, we also test performance on fission reactions outside the training set: 
neutron-induced fission on $^{233}$U, $^{234}$U, $^{237}$Np, and $^{241}$Pu. In Fig. \ref{fig:delta_nubar_multi}(b),
we see that $\Delta \bar \nu / \bar \nu$ is at most a factor of 2 greater than that for the trained reactions. 
The largest percent error is seen for $^{237}$Np(n,f), which did not have any reaction with the same 
compound charge included in the training set.  Still, the relative error is within or under the experimental uncertainty on \nup{}.
Across all nuclei there is a 
consistent upward trend in $\Delta \bar \nu / \bar \nu$ as incident neutron energy increases. This discrepancy is due to slightly 
worse performance of the emulator for higher $\nu$ values, as seen in Fig. \ref{fig:cf252_dist}(b). We infer that this is caused by 
the more frequent incidence of high multiplicities at higher incident neutron energies, which comprise a small 
fraction of the overall training set. Despite this challenge, worst performance on reactions outside the training set is 
around 0.6\%; this small discrepancy gives us confidence 
that the emulator is capable of generalizing to new reactions and has learned a useful mapping from fission fragment initial conditions  
to multiplicities for prompt neutrons.

\subsection{$\gamma$-ray multiplicity}
\begin{figure}
    \includegraphics[width=0.45\textwidth]{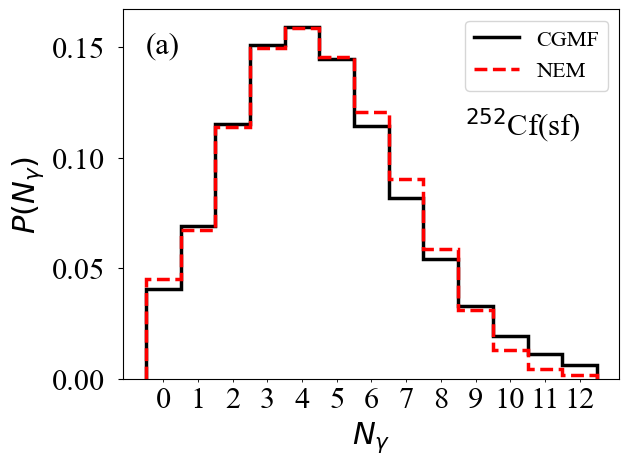}
    \includegraphics[width=0.45\textwidth]{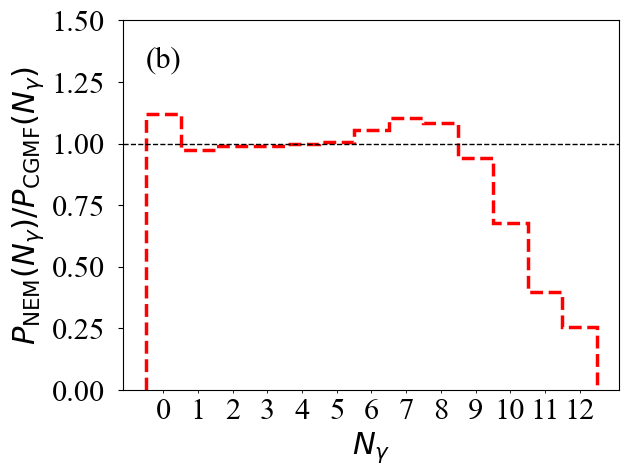}
    \caption{(a) Comparison of the \gray{} multiplicity distribution for $^{252}$Cf between \cgmf{} and the emulator. 
    While performance is worse than for neutrons, the emulator still reproduces a value $\bar{\hat{N}}_\gamma= 4.41$ within 2\% of \cgmf{}'s 
    value, $\bar{N}_\gamma = 4.50$.
  (b) Ratio of the prediction to the empirical value.
}
    \label{fig:cf252_dist_gamma}
\end{figure}
Figure \ref{fig:cf252_dist_gamma} shows the emulator's performance for \grayd{} multiplicity, again using the example of $^{252}$Cf spontaneous fission. 
We observe that the quality of the \grayd{} multiplicity NEM is significantly worse 
than the neutron multiplicity emulator, particularly in the high multiplicity (low incidence) regime. However,
performance in the most common multiplicities (0-9) remains reliable to within 12\%.

\begin{figure*}
    \includegraphics[width=0.45\textwidth]{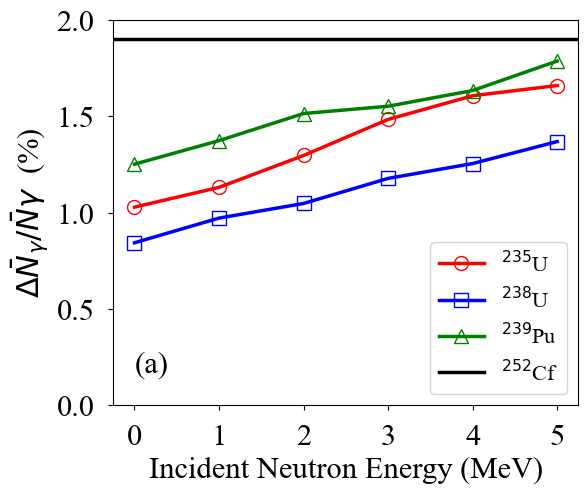}
    \includegraphics[width=0.45\textwidth]{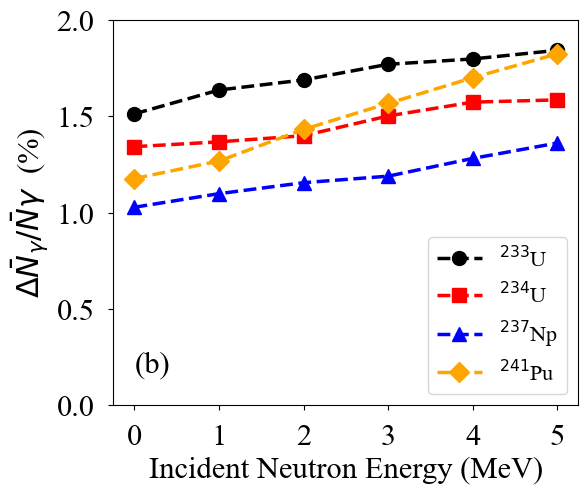}
    \caption{(a) Percent error in the average \gray{} multiplicity, $\bar{N}_\gamma$, between \cgmf{} and the emulator, as in Eq. (\ref{eqn:delta}), 
    for $^{235}$U(n,f) (red), $^{238}$U(n,f) (blue), and $^{239}$Pu(n,f) (green) as a function of incident neutron energy. Error for spontaneous fission 
    of $^{252}$Cf(n,f) is shown in black.
    (b) Same as (a) except for reactions that were not included in the training set: $^{233}$U(n,f) (black), $^{234}$U(n,f) (red), $^{237}$Np(n,f) (blue), and $^{241}$Pu(n,f) (orange).
}
    \label{fig:delta_gammabar_multi}
\end{figure*}

When viewed at the reaction level in Fig. \ref{fig:delta_gammabar_multi}(a), which shows the percent error difference between the NEM and \cgmf{}, the error of $\hat \gamma$ is still 
quite low for trained reactions, with $\Delta \bar N_\gamma / \bar N_\gamma$ peaking below 2\% (note that $\Delta \bar N_\gamma / \bar N_\gamma$ is 
defined the same way as in Eq. (\ref{eqn:delta}) with \nup{} replaced with \nug{}). Similarly, $\hat \gamma$ generalizes well to new reactions. 
$\Delta \bar N_\gamma / \bar N_\gamma$ ranges from about 1\% to 2\% for the four fission reactions from outside the training
set, as in Fig. \ref{fig:delta_gammabar_multi} (b). We note the 
same trend of increasing error with increasing incident neutron energy as was seen for the neutron NEM.

\subsection{Discussion}
General performance for the neutron multiplicity emulator is excellent, with $\Delta \bar \nu / \bar \nu$ falling below
0.4\% for trained reactions and 0.6\% on reactions outside the training set. This error is on the order of or lower than the deviation 
of \cgmf{} from experimental data, making this emulator promising as a rapid fission event generator.

The neutron NEM's performance on individual fission events is what gives it the generality
to model whole fission reactions, including those outside the training set, with such fidelity. One primary 
use case for this emulator is uncertainty quantification (UQ) of \cgmf{} itself. Performing UQ requires the adjustment of 
many parameters within \cgmf{} and iterative testing of how these adjustments change the outputs. If we were to 
test this emulator only on fission reactions in the training data, it would not be clear whether the emulator 
would remain valid under adjustments to \cgmf{}'s parameters, because these adjustments lead to different 
distributions for the post-scission conditions. However, by testing our emulator on fission reactions outside 
the training set, we gain confidence that the emulator remains accurate even when \cgmf{} parameters are changed.

The \grayd{} NEM, $\hat \gamma$, has worse performance than the neutron emulator $\hat \nu$, with $\Delta \bar N_\gamma / \bar N_\gamma$ on trained 
reactions rising about an order of magnitude. 
However, for the ultimate goal of performing UQ with this \cgmf{} emulator, we note that experimental measurements 
of \nup{} can be at the 0.5--1\% level where as measurements of \nug{} usually have an uncertainty of a few percent.  
Therefore, both of these NEM emulators are within the tolerance of the experimental measurements and should not 
bias any optimization that can be performed.  

\section{Results for energy emulation}
\label{sec:resultsEnergy}

\begin{figure}
\centering
\includegraphics[width=0.5\textwidth]{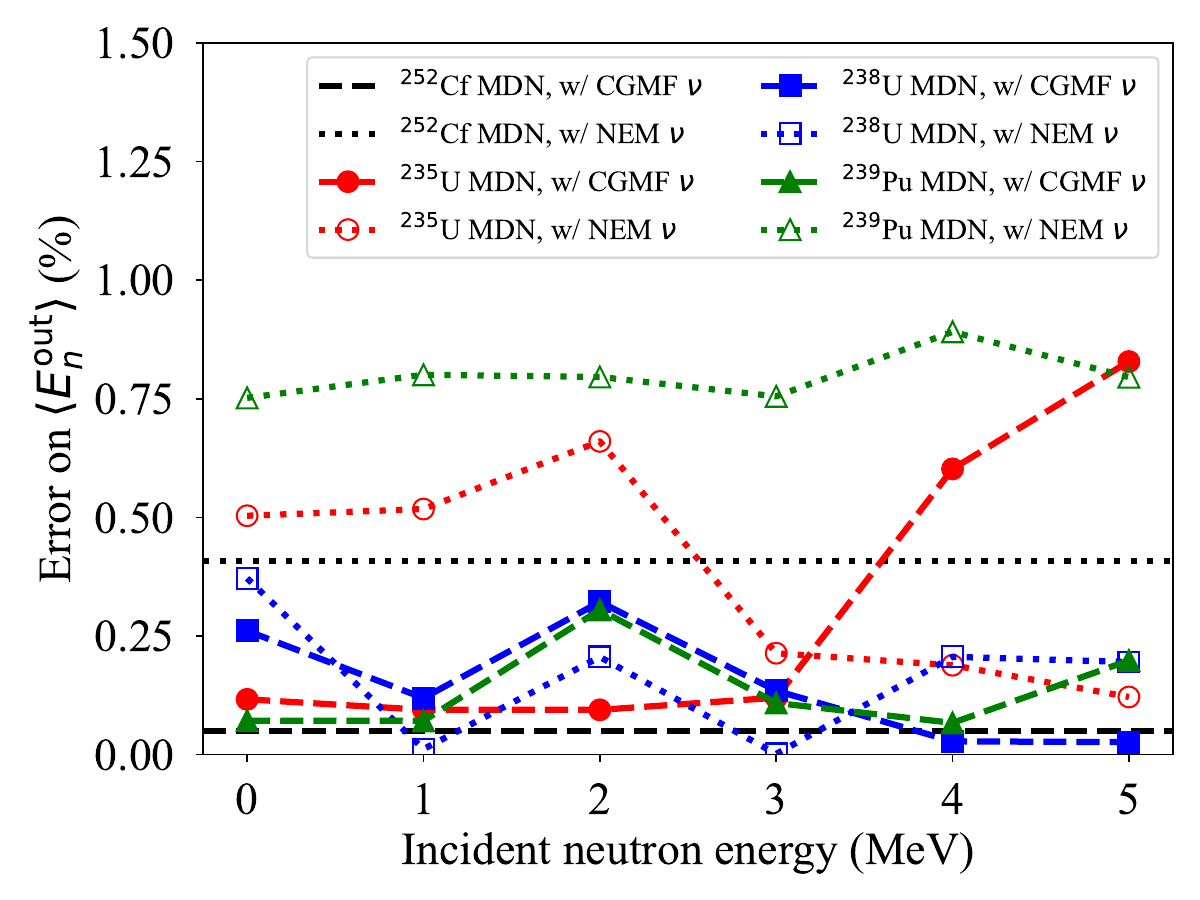}
\caption{Percent error of the average outgoing neutron energies, $\langle E^\mathrm{out}_n \rangle$, as a function of incident neutron energy for $^{235}$U (red circles), $^{238}$U (blue squares), and $^{239}$Pu (green triangles), compared to the full \cgmf{} calculation.  Dashed lines and filled symbols indicate the neutron MDN using neutron multiplicities from \cgmf{}; dotted lines with open symbols use the neutron multiplicity from the NEM.  Comparisons for the spontaneous fission of $^{252}$Cf are shown by the dashed black line (MDN with \cgmf{} multiplicities) and dotted black line (MDN with emulator multiplicities).}
\label{fig:neutronEnergies}
\end{figure}

We next show results for the neutron and \grayd{} MDNs for energy emulation.  Figure \ref{fig:neutronEnergies} depicts 
the percent error on the average neutron energies between \cgmf{} and the 
neutron MDN for all reactions included 
in the training set as a function of the incident neutron energy.  The percent error on $^{252}$Cf spontaneous 
fission is shown as a horizontal line across the whole incident energy range.

For the neutron MDN, we compare two calculations. In the first, we use the neutron 
multiplicities calculated directly from \cgmf{} (filled symbols), and for the second, we use the multiplicities 
from the neutron NEM (open symbols, described in Sec. \ref{sec:global}).  
When the multiplicities from \cgmf{} are used directly in the energy emulator, the agreement between the emulator
and \cgmf{} is better than 0.3\% for all incident neutron energies.  When the multiplicities from the 
NEM are included into the MDN, the percent error is larger, but still below 1\% for all reactions 
and incident energies in the training set.  The largest discrepancies are seen for $^{239}$Pu, likely due to the 
higher average neutron multiplicity across the whole energy range.  Although the percent errors on \nup{} are 
below 0.5\% for most 
reactions, the event-by-event differences in the multiplicities have a larger impact on the average neutron energies.

\begin{figure}
\centering
\includegraphics[width=0.5\textwidth]{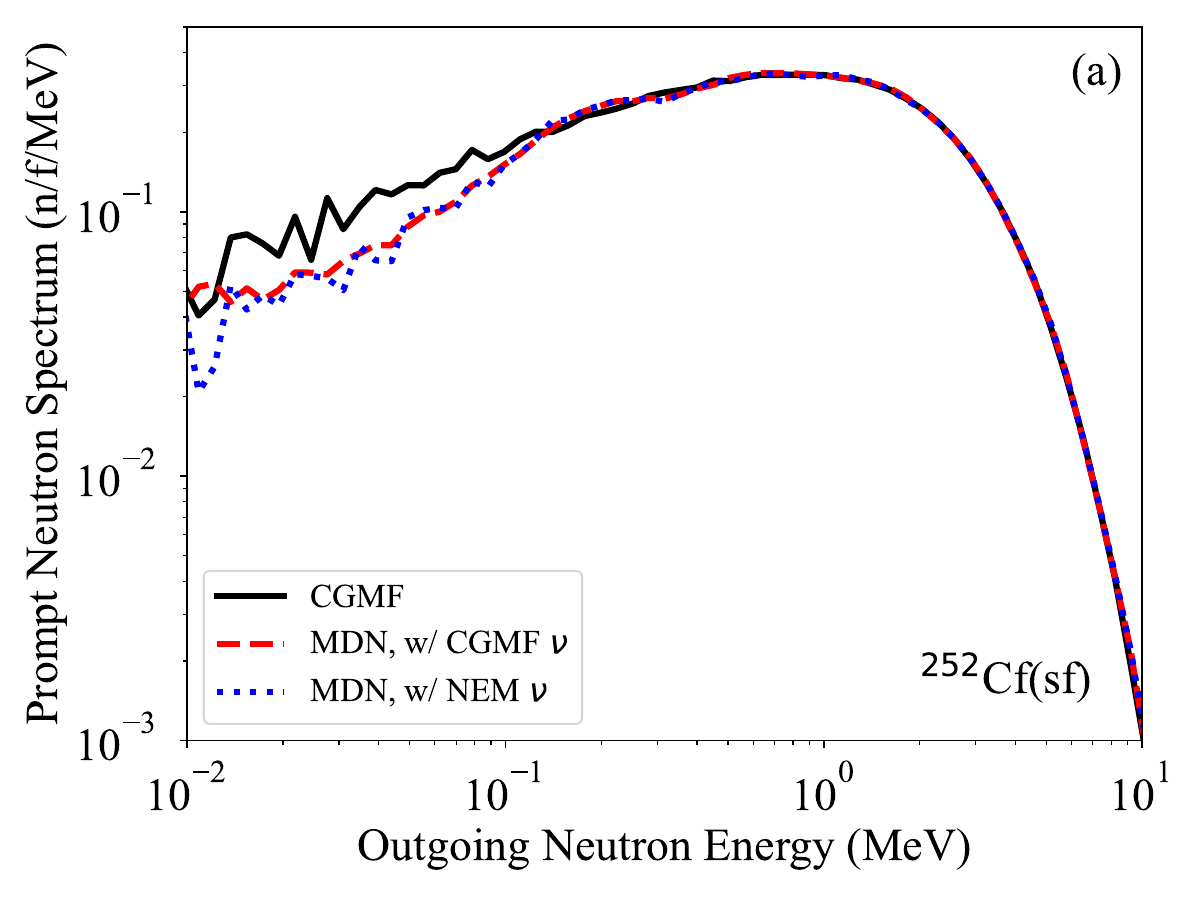} \\
\includegraphics[width=0.5\textwidth]{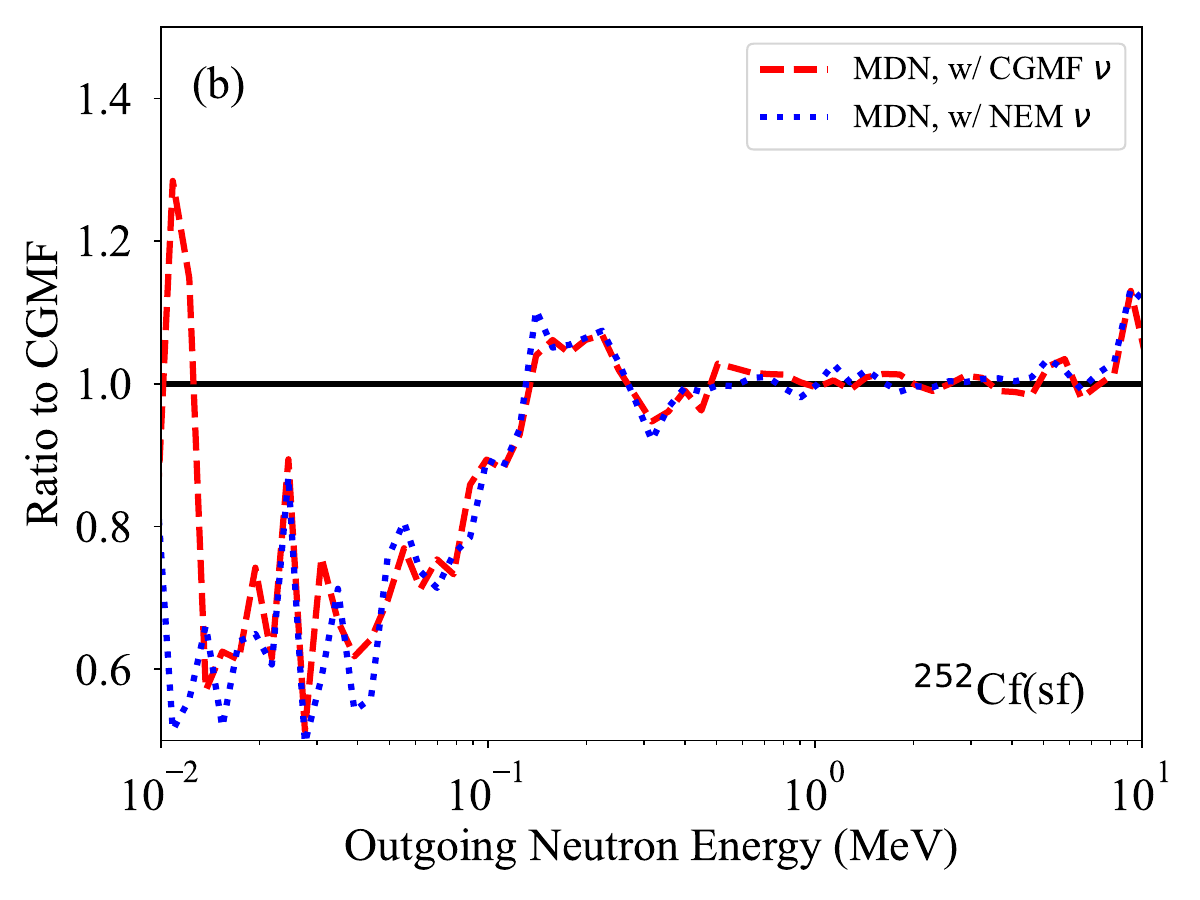} \\
\caption{(a) Prompt fission neutron spectrum, PFNS, as a function of outgoing neutron energies, for the spontaneous fission of $^{252}$Cf.  (b) Ratio of the $^{252}$Cf(sf) PFNS using the MDN emulator to the PFNS calculated from \cgmf{}.  In both panels, the \cgmf{} calculation is shown by the solid black line and the MDN with neutron multiplicities from \cgmf{} (the NEM) by the red dashed (blue dotted) line.}
\label{fig:PFNS}
\end{figure}

In Fig. \ref{fig:PFNS}, we show a comparison of the prompt fission neutron 
spectrum (PFNS), again comparing \cgmf{} 
(black solid), the MDN prediction using \cgmf{} neutron multiplicities (red dashed), and the 
MDN prediction using the NEM neutron multiplicities (blue dotted).  Panel (a) shows the comparison of the 
absolute PFNS, while panel (b) shows the comparison as a ratio to the \cgmf{} PFNS.  The largest differences
in the PFNS are seen when the outgoing neutron energies are below about 100 keV.  Again, we see little difference between the MDN results when multiplicities 
are used from \cgmf{} compared to the noisy emission model.  The average difference between \cgmf{} and either 
of the emulators is better than 3\%.  

\begin{figure}
\centering
\includegraphics[width=0.5\textwidth]{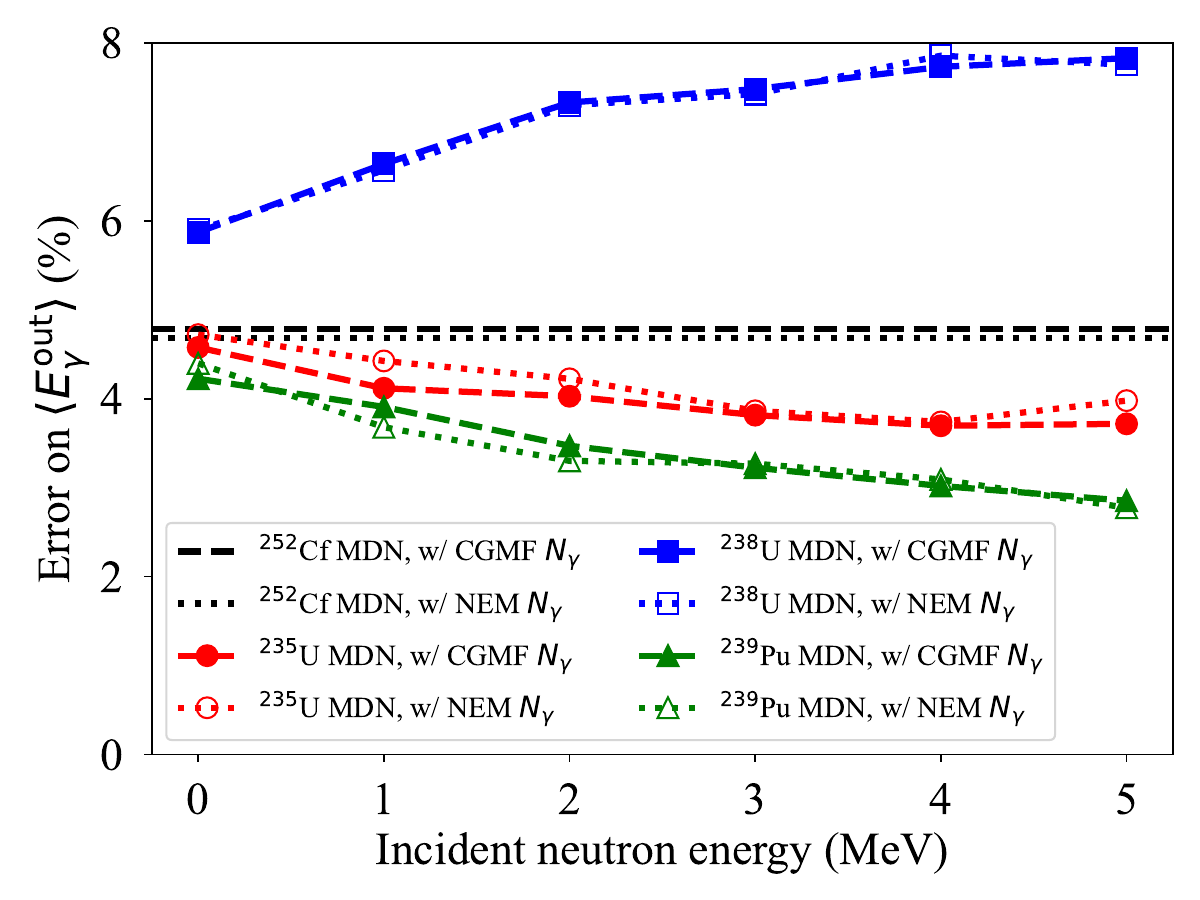}
\caption{Same as Fig. \ref{fig:neutronEnergies} for the average outgoing \grayd{} energies, $\langle E^\mathrm{out}_\gamma \rangle$.}
\label{fig:gammaEnergies}
\end{figure}

In Fig. \ref{fig:gammaEnergies}, we show the percent error on the average \grayd{} energy.  
Here, we see that the average energies do not follow the trend of \cgmf{} across incident neutron energy, neither when
 using multiplicities from \cgmf{} (filled symbols) or from the \grayd{} emulator 
(open symbols).  Unlike for the neutron energies, the MDN cannot distinguish between the 
different compound nuclei that are fissioning, and the average \grayd{} energies are essentially 
flat as a function of incident energy and target.  This feature is unsurprising considering 
that the fission fragment initial conditions are a poor indicator of which fission fragments are 
actually emitting \gray{}s and the amount of excitation energy remaining after the neutrons are 
emitted.  The neutrons each remove around 1 MeV plus the one-neutron separation energy 
of excitation energy from the decaying fission fragment, leaving significantly less energy for the \gray{}s. Additionally, we note that in \cgmf{}, 
the neutrons are able to carry away significant angular momentum in \cgmf{}, much more than 
0.5$\hbar$ \cite{Stetcu2021}.  While it has been shown that putting more physics-based 
information into the input of a NN training set, e.g. \cite{Lovell2022}, can improve predictions, attempts to train the 
\grayd{} MDN with the mass of the compound nucleus after neutron emission ($A-\nu$) 
and $U$ minus the energy of the emitted neutrons did not perform any 
better than the results shown here.  

\begin{figure}
\centering
\includegraphics[width=0.5\textwidth]{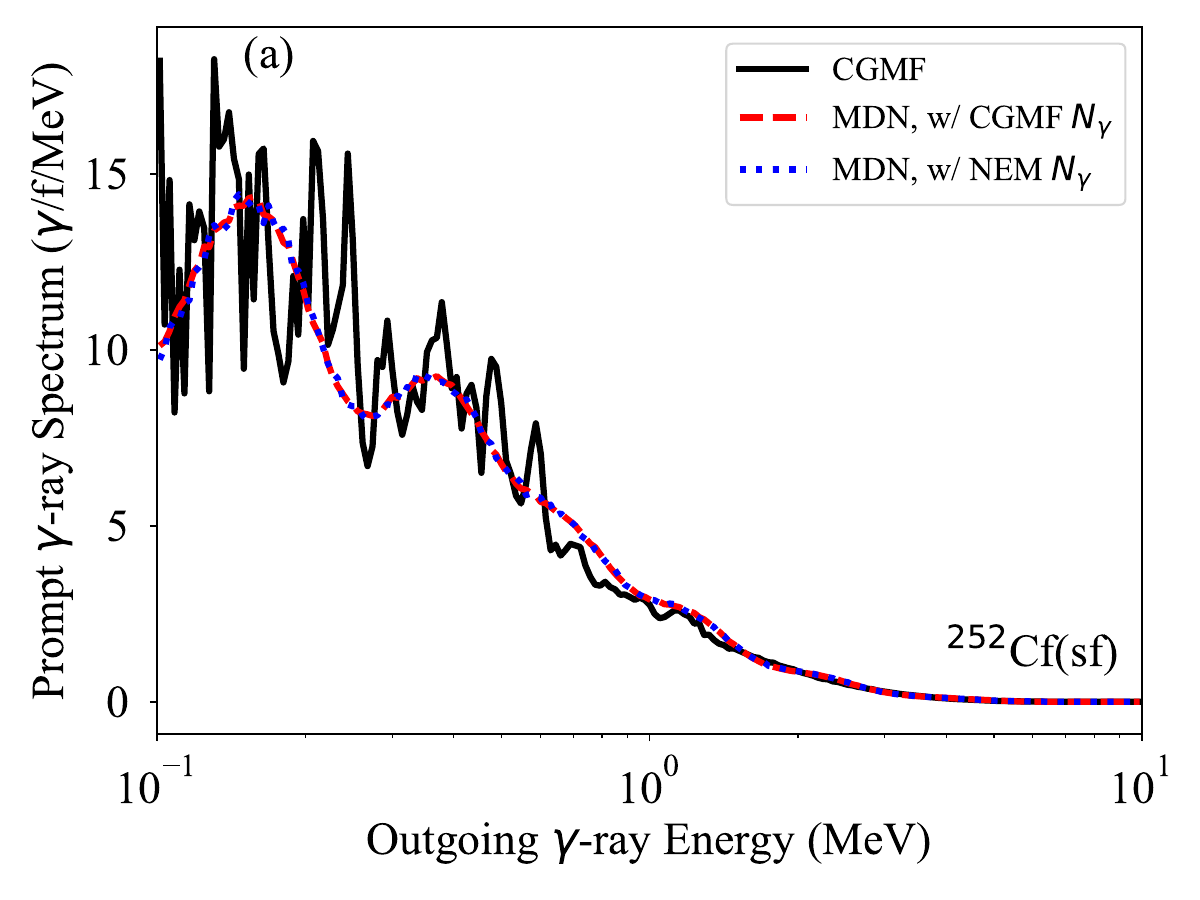} \\
\includegraphics[width=0.5\textwidth]{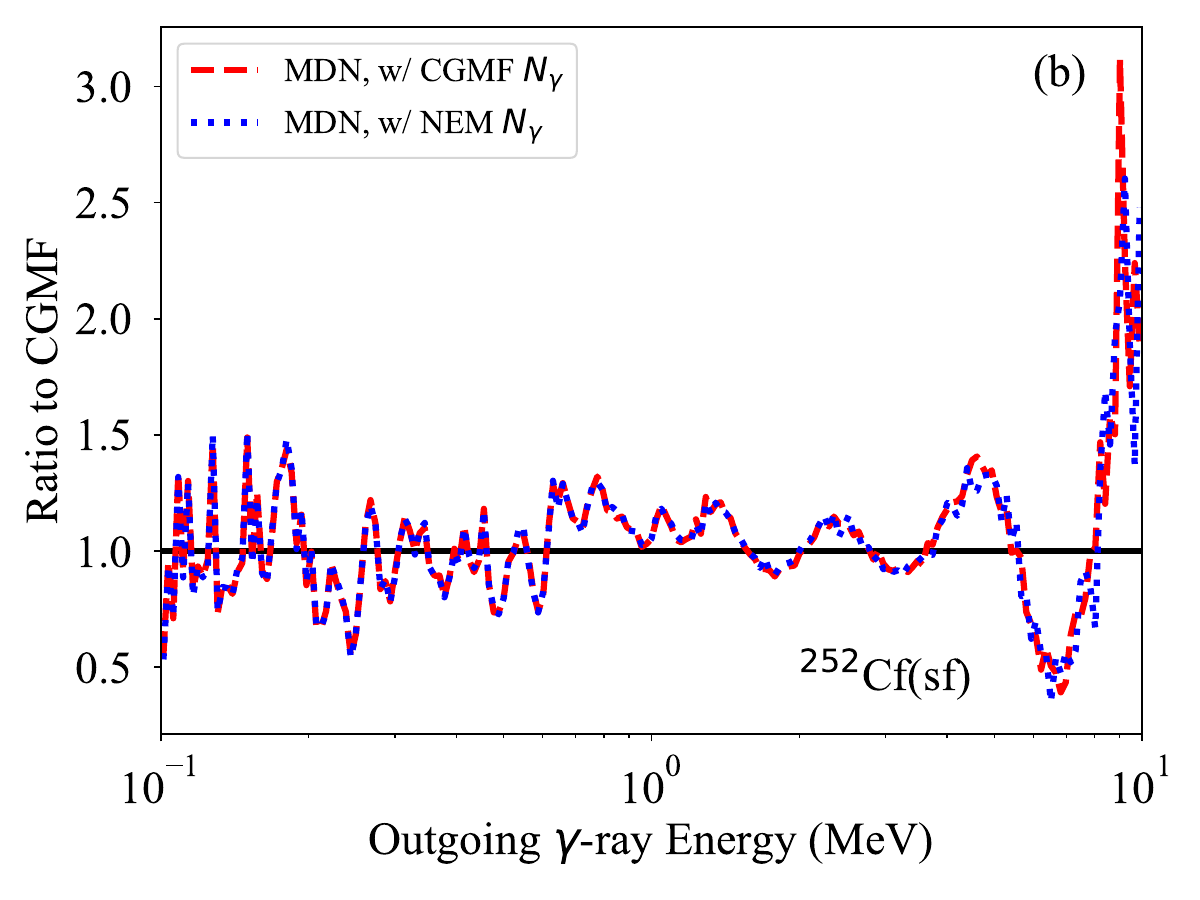} \\
\caption{Same as Fig. \ref{fig:PFNS} for the prompt fission \grayd{} spectrum, PFGS.}
\label{fig:PFGS}
\end{figure}

The prompt fission \grayd{} 
spectrum, PFGS, is plotted in Fig. \ref{fig:PFGS}(a), illustrating why the average \grayd{} energies are discrepant 
from \cgmf{}.  Although the emulator reproduces the tail of the PFGS, it cannot reproduce the discrete \grayd{} lines that are known from nuclear structure.  We can reproduce this 
structure somewhat more exactly, following more of the peaks below 2 MeV, when we train on 
only a single target nucleus.  This discrepancy again indicates that the fission fragment initial conditions 
including the \grayd{} MDN are not sufficient to distinguish between target nuclei.  We do note, however, 
that the tail of the spectrum is better reproduced than the discrete \grayd{} peaks, indicated in 
Fig. \ref{fig:PFGS}(b).  This agreement indicates that the \grayd{} emulator is at least learning the trend between 
the fission fragment spin distribution and the PFGS.  

\section{Full emulator}
\label{sec:fullEmulator}

We finally discuss the importance of the full emulator for fission studies with \cgmf{}.  The two NEMs and 
two MDNs have been interfaced with the model of \cgmf{} that only calculates the fission fragment initial 
conditions.  We find about three orders of magnitude speed up between the full \cgmf{} calculation and 
the emulator described in this work.  Currently, the speed of our emulator is limited by the time needed 
to sample the fission fragment initial conditions, which has recently been improved \cite{Talou2023} compared to \cite{CGMF} for 
the incident energy range considered here.

To show the power of the full emulator, we construct several models for the spontaneous fission of $^{252}$Cf 
that differ by varying a handful of parameters that determine the fission fragment initial conditions.  Namely, 
we vary the average total kinetic energy, $\overline{\mathrm{TKE}}$, the width of the TKE distribution, 
$\sigma_\mathrm{TKE}$, and the spin cutoff parameter, $\alpha$.  The default values from \cgmf{} \cite{CGMF} 
and the updated values are shown in the second, third, and fourth columns of Table \ref{tab:models}.  We 
have chosen these three quantities to vary because previous studies have shown their impact on prompt 
fission observables.  Several previous studies have demonstrated that $\overline{\mathrm{TKE}}$ and \nup{} are strongly anti-correlated 
\cite{Jaffke2018,Randrup2019}.  The width of the TKE distribution has an impact on the width of the neutron 
multiplicity distribution \cite{Talou2018}, often denoted as the second moment of the distribution, $\nu_2$.  
The spin cutoff parameter has a relatively small impact on the neutron observables but does strongly impact 
the \grayd{} properties~\cite{Stetcu2014}.  

\begin{table*}
\centering
\begin{tabular}{c|ccc|ccccc}
\textbf{Model name} & $\overline{\mathrm{TKE}}$ (MeV) & $\sigma _\mathrm{TKE}$ (MeV) & $\alpha$ & \nup{} & $\nu_2$ & \nug{} & $\langle E^\mathrm{out}_n \rangle $ (MeV) & $\langle E^\mathrm{out}_\gamma \rangle $ (MeV)\\ \hline \hline
\cgmf{} & 185.78 & 7.3846 & 1.7 & 3.814 & 11.852 & 8.709 & 2.085 & 0.760 \\
M1 & 184.78 & 7.3846 & 1.7 & 3.949 & 12.737 & 8.713 & 2.102 & 0.761 \\
M2 & 186.78 & 7.3846 & 1.7 & 3.680 & 11.011 & 8.702 & 2.078 & 0.760 \\
M3 & 185.78 & 8.3846 & 1.7 & 3.805 & 12.035 & 8.719 & 2.090 & 0.760 \\
M4 & 185.78 & 7.3846 & 1.5 & 3.848 & 12.067 & 8.355 & 2.086 & 0.760 \\
\end{tabular}
\caption{Average total kinetic energy (in MeV), width of the total kinetic energy distribution, and the spin 
cutoff parameter (second, third, and fourth columns respectively), for each of the models run through the 
full emulator for the spontaneous fission of $^{252}$Cf.  The model denoted as \cgmf{} indicates the baseline values 
from \cgmf{} \cite{CGMF}.  Columns five, six, seven, eight, and nine give the average prompt neutron multiplicity, 
second moment of the neutron multiplicity distribution, average prompt \grayd{} multiplicity, average outgoing 
neutron energy, and average outgoing \grayd{} energy that result from these models being run using the full emulator.}
\label{tab:models}
\end{table*}

In columns five through nine in Table \ref{tab:models}, we show the results from the prompt fission 
observables described above to changes in these three inputs, using the full emulator.  From models 
$M1$ and $M2$, which lower and raise the $\overline{\mathrm{TKE}}$ by 1 MeV, \nup{} is increased or decreased by 0.135 neutrons respectively.  With our emulator, the strong anti-correlation between 
these two quantities is recovered, and the relative changes are similar in value to what was shown in \cite{Randrup2019}.  
The correlations between $\overline{\mathrm{TKE}}$ and the average outgoing neutron energy, $\langle E_n^\mathrm{out} \rangle$, can appear to be contradictory at first glance.  When $\overline{\mathrm{TKE}}$ decreases, $\langle E_n^\mathrm{out} \rangle$ increases; even though there is 
less kinetic energy to boost the neutrons in the laboratory frame (see e.g. \cite{CGMF,Lovell2020a} for a 
discussion of the kinematics of the fission fragments and neutrons), the energy of the neutrons increase 
on average because of the additional excitation energy available for the decay.  The opposite trend is 
observed when $\overline{\mathrm{TKE}}$ is increased.  
When we increase $\sigma_\mathrm{TKE}$ using $M3$, we also see the expected increase in $\nu_2$, which 
is defined \cite{Talou2018} as
\begin{equation}
\nu_n = \sum _\nu \frac{\nu!}{(\nu-n)!} P(\nu),
\end{equation}
for $n=2$.  However, we note that there is a clear impact of both $\overline{\mathrm{TKE}}$ and 
$\sigma_\mathrm{TKE}$ on \nup{} and $\nu_2$.  

Finally, changing $\alpha$ has a smaller, but non-negligible, impact on \nup{}.  However, there are larger 
($\sim 4$\%) changes on \nug{}.  Although these changes are larger than what was calculated in \cite{Stetcu2014}, 
we show that our emulator reproduces anti-correlations between the neutron and \grayd{} multiplicities 
when the spin distribution is changed.  We additionally note that even though $\alpha$ should change the 
average \grayd{} energy, $\langle E^\mathrm{out}_\gamma \rangle $, we see no change when using our 
emulator.  For all of the input changes, only a change of 1 keV in $\langle E^\mathrm{out}_\gamma \rangle $ 
is seen, consistent with the \grayd{} MDN not being able to distinguish energies based on fission fragment 
initial conditions.  For broader applicability of our emulator, the \grayd{} energies will need significant improvement.

\section{Conclusion}
\label{sec:conclusion}

Here, we produce an emulator for event-by-event fission fragment decay code, \cgmf{}, using a 
combination of a noisy emission models (NEMs) and a mixture density networks (MDNs).  The NEMs 
are used to emulate the multiplicity of the neutrons and \gray{}s from a fission fragment, given a 
subset of the fission fragment initial conditions in mass, charge, kinetic energy, excitation energy, 
spin, and parity.  The MDNs emulate the energies of each neutron or \gray{} emitted, given the 
multiplicities predicted by the NEM.  The neutron emulators can reproduce the full \cgmf{} results 
to within a percent or better for both neutron energies and multiplicities, while the \grayd{} emulator 
suffers from larger discrepancies on the order of a few percent.  The emulator generalizes to 
reactions outside of the training set---which included spontaneous fission of $^{252}$Cf and 
neutron-induced fission up to 5 MeV for $^{235}$U, $^{238}$U, and $^{239}$Pu---to an error 
within a factor of two compared to the training set.  We find a speed up of about three orders 
of magnitude using our emulator compared to a full \cgmf{} calculation, which is mainly limited 
by the calculation of the fission fragment initial conditions, not by the implementation of the emulator.  

While providing impressive accuracy and speed gains, our emulator has several limitations that require further investigation. The \grayd{} emulator in
particular falls short of the accuracy achieved by the neutron emulator, due to the more complex distributions
to be learned and the difference between fission fragment initial conditions and the fragment conditions when 
the \gray{}s are emitted.
Another limitation is event-by-event performance, particularly for the NEM. Although, for example, the neutron multiplicity 
distribution from the neutron NEM reproduces \cgmf{} to better than 1\% on average, when these multiplicities are used as inputs to the neutron MDN, the discrepancy between the neutron energies from the emulator and \cgmf{} worsens.  

However, the results shown in this paper open up one avenue into emulating event generators, rather 
than the typical structure and reaction observables that have already been the focus of a broad development 
of emulators for nuclear theory models.  The pathway sets the stage for future studies into robust uncertainty 
quantification (UQ) of the fission fragment initial conditions, following work such as \cite{Jaffke2018,Randrup2019}.  
Such robust UQ studies will allow us to begin to disentangle whether discrepancies in the full model are coming 
from initial conditions, global models for structure properties, or model deficiencies.  

Future work includes further emulator development, particularly looking at machine learning methods that can 
emulate the partial differential equations that are solved within \cgmf{}, rather than constructing a blackbox 
mapping from inputs to outputs.  Similar work has been the focus of emulators such as \cite{Odell2024} which 
builds an emulator for neutron scattering but has not yet been implemented into \cgmf{} or other event generators.  
The goal is to be able to provide a similar speed up to the current emulator with better accuracy for both neutron 
and \grayd{} observables.  Additionally, although the $\texttt{pytorch}$ implementation of this emulator provides a 
significant speed up, the separate call outside of \cgmf{} could limit the emulator's use in transport codes, such as MCNP\textsuperscript{\textregistered}~\cite{MCNP}.  Work to translate the emulator into C\texttt{++} is planned to more directly integrate it with \cgmf{}.

\begin{acknowledgements}
We acknowledge funding from the U.S. Department of Energy Laboratory Directed Research and Development
 program and support from the XCP Computation Physics Workshop, both at Los Alamos National Laboratory. 
This work was performed under the auspice of the U.S. Department of Energy by Los Alamos National Laboratory under Contract 89233218CNA000001.  
ROS acknowledges support from NSF PHY 2012057, PHY 2309172, AST 2206321, and the Simons Foundation.
\end{acknowledgements}

\newpage

\appendix

\section{Event-by-event distribution estimation}

\label{ap:event}
For our model to be useful and general enough for its intended uses, we require it to be capable
of representing distributions for multiplicities on an event-by-event basis. Concretely, the model 
must reproduce distributions $\hat\nu(\mathbf{x})$ based on the true distribution $\nu(\mathbf{x})$ at any given point $\mathbf{x}$ in the
parameter space. However, as discussed in Sec. \ref{sec:design}, we do not have direct access to these
distributions $\nu(\mathbf{x})$; we only have discrete samples taken from the underlying distribution at various points
across parameter space. 

To allow us to validate our model on an event-by-event basis, we use a strategy we will refer to as local clustering.
In principle, if we group together many data points centered around
a given point $\mathbf{x}$ in parameter space, we can obtain an estimate of the true distribution $\nu(\mathbf{x})$ by counting the 
relative frequency of multiplicity outcomes (0, 1, 2, ...) in that local cluster. By randomly selecting many points $\mathbf{x}_v$
within the full dataset and clustering the nearby datapoints $\mathbf{x}_v^c$ of each, we can construct an estimate for $\nu$
by assuming $\nu(\mathbf{x}_v)$ maps to the distribution defined by the outcomes for multiplicity associated with 
$\mathbf{x}_v^c$. Having estimated $\nu$, we can validate $\hat{\nu}$ with $\Delta \bar \nu$, defined by Eq. (\ref{eqn:delta}).

For summarizing the performance of the model across an entire data set, e.g. for the decay reactions
created by a specific fission event such as $^{252}$Cf, it is useful to evaluate the model on an entire dataset
and tally the results. These results form a distribution of multiplicities resulting from a given fission
event and can be compared directly to the distribution observed in the dataset. While such a method does
not prove event-by-event performance, it is a useful metric for many of the possible applications of 
the emulator.

Validation datasets were constructed using the method described in Sec. \ref{sec:validate}. To
build the validation set, a separate run of \cgmf{} was used with 1,000,000 data points per fission event
for a total of 19 million points. 

\bibliography{references.bib}

\begin{thebibliography}{54}
\expandafter\ifx\csname natexlab\endcsname\relax\def\natexlab#1{#1}\fi
\expandafter\ifx\csname bibnamefont\endcsname\relax
  \def\bibnamefont#1{#1}\fi
\expandafter\ifx\csname bibfnamefont\endcsname\relax
  \def\bibfnamefont#1{#1}\fi
\expandafter\ifx\csname citenamefont\endcsname\relax
  \def\citenamefont#1{#1}\fi
\expandafter\ifx\csname url\endcsname\relax
  \def\url#1{\texttt{#1}}\fi
\expandafter\ifx\csname urlprefix\endcsname\relax\def\urlprefix{URL }\fi
\providecommand{\bibinfo}[2]{#2}
\providecommand{\eprint}[2][]{\url{#2}}

\bibitem[{\citenamefont{Jaffke et~al.}(2018)\citenamefont{Jaffke, M\"oller,
  Talou, and Sierk}}]{Jaffke2018}
\bibinfo{author}{\bibfnamefont{P.}~\bibnamefont{Jaffke}},
  \bibinfo{author}{\bibfnamefont{P.}~\bibnamefont{M\"oller}},
  \bibinfo{author}{\bibfnamefont{P.}~\bibnamefont{Talou}}, \bibnamefont{and}
  \bibinfo{author}{\bibfnamefont{A.~J.} \bibnamefont{Sierk}},
  \bibinfo{journal}{Phys. Rev. C} \textbf{\bibinfo{volume}{97}},
  \bibinfo{pages}{034608} (\bibinfo{year}{2018}),
  \urlprefix\url{https://link.aps.org/doi/10.1103/PhysRevC.97.034608}.

\bibitem[{\citenamefont{Randrup et~al.}(2019)\citenamefont{Randrup, Talou, and
  Vogt}}]{Randrup2019}
\bibinfo{author}{\bibfnamefont{J.}~\bibnamefont{Randrup}},
  \bibinfo{author}{\bibfnamefont{P.}~\bibnamefont{Talou}}, \bibnamefont{and}
  \bibinfo{author}{\bibfnamefont{R.}~\bibnamefont{Vogt}},
  \bibinfo{journal}{Phys. Rev. C} \textbf{\bibinfo{volume}{99}},
  \bibinfo{pages}{054619} (\bibinfo{year}{2019}),
  \urlprefix\url{https://link.aps.org/doi/10.1103/PhysRevC.99.054619}.

\bibitem[{\citenamefont{Talou et~al.}(2018)\citenamefont{Talou, Vogt, Randrup,
  Rising, Pozzi, Verbeke, Andrews, Clarke, Jaffke, Jandel et~al.}}]{Talou2018}
\bibinfo{author}{\bibfnamefont{P.}~\bibnamefont{Talou}},
  \bibinfo{author}{\bibfnamefont{R.}~\bibnamefont{Vogt}},
  \bibinfo{author}{\bibfnamefont{J.}~\bibnamefont{Randrup}},
  \bibinfo{author}{\bibfnamefont{M.~E.} \bibnamefont{Rising}},
  \bibinfo{author}{\bibfnamefont{S.~A.} \bibnamefont{Pozzi}},
  \bibinfo{author}{\bibfnamefont{J.}~\bibnamefont{Verbeke}},
  \bibinfo{author}{\bibfnamefont{M.~T.} \bibnamefont{Andrews}},
  \bibinfo{author}{\bibfnamefont{S.~D.} \bibnamefont{Clarke}},
  \bibinfo{author}{\bibfnamefont{P.}~\bibnamefont{Jaffke}},
  \bibinfo{author}{\bibfnamefont{M.}~\bibnamefont{Jandel}},
  \bibnamefont{et~al.}, \bibinfo{journal}{The European Physical Journal A}
  \textbf{\bibinfo{volume}{54}}, \bibinfo{pages}{9} (\bibinfo{year}{2018}),
  \urlprefix\url{https://doi.org/10.1140/epja/i2018-12455-0}.

\bibitem[{\citenamefont{Schunck et~al.}(2014)\citenamefont{Schunck, Duke, Carr,
  and Knoll}}]{Schunck2014}
\bibinfo{author}{\bibfnamefont{N.}~\bibnamefont{Schunck}},
  \bibinfo{author}{\bibfnamefont{D.}~\bibnamefont{Duke}},
  \bibinfo{author}{\bibfnamefont{H.}~\bibnamefont{Carr}}, \bibnamefont{and}
  \bibinfo{author}{\bibfnamefont{A.}~\bibnamefont{Knoll}},
  \bibinfo{journal}{Phys. Rev. C} \textbf{\bibinfo{volume}{90}},
  \bibinfo{pages}{054305} (\bibinfo{year}{2014}),
  \urlprefix\url{https://link.aps.org/doi/10.1103/PhysRevC.90.054305}.

\bibitem[{\citenamefont{Sierk}(2017)}]{Sierk2017}
\bibinfo{author}{\bibfnamefont{A.~J.} \bibnamefont{Sierk}},
  \bibinfo{journal}{Phys. Rev. C} \textbf{\bibinfo{volume}{96}},
  \bibinfo{pages}{034603} (\bibinfo{year}{2017}),
  \urlprefix\url{https://link.aps.org/doi/10.1103/PhysRevC.96.034603}.

\bibitem[{\citenamefont{Verriere et~al.}(2019)\citenamefont{Verriere, Schunck,
  and Kawano}}]{Verriere2019}
\bibinfo{author}{\bibfnamefont{M.}~\bibnamefont{Verriere}},
  \bibinfo{author}{\bibfnamefont{N.}~\bibnamefont{Schunck}}, \bibnamefont{and}
  \bibinfo{author}{\bibfnamefont{T.}~\bibnamefont{Kawano}},
  \bibinfo{journal}{Phys. Rev. C} \textbf{\bibinfo{volume}{100}},
  \bibinfo{pages}{024612} (\bibinfo{year}{2019}),
  \urlprefix\url{https://link.aps.org/doi/10.1103/PhysRevC.100.024612}.

\bibitem[{\citenamefont{Regnier et~al.}(2016)\citenamefont{Regnier, Dubray,
  Schunck, and Verri\`ere}}]{Regnier2016}
\bibinfo{author}{\bibfnamefont{D.}~\bibnamefont{Regnier}},
  \bibinfo{author}{\bibfnamefont{N.}~\bibnamefont{Dubray}},
  \bibinfo{author}{\bibfnamefont{N.}~\bibnamefont{Schunck}}, \bibnamefont{and}
  \bibinfo{author}{\bibfnamefont{M.}~\bibnamefont{Verri\`ere}},
  \bibinfo{journal}{Phys. Rev. C} \textbf{\bibinfo{volume}{93}},
  \bibinfo{pages}{054611} (\bibinfo{year}{2016}),
  \urlprefix\url{https://link.aps.org/doi/10.1103/PhysRevC.93.054611}.

\bibitem[{\citenamefont{M\"oller and Ichikawa}(2015)}]{Moller2015}
\bibinfo{author}{\bibfnamefont{P.}~\bibnamefont{M\"oller}} \bibnamefont{and}
  \bibinfo{author}{\bibfnamefont{T.}~\bibnamefont{Ichikawa}},
  \bibinfo{journal}{Eur. Phys. J. A} \textbf{\bibinfo{volume}{51}},
  \bibinfo{pages}{173} (\bibinfo{year}{2015}),
  \urlprefix\url{https://doi.org/10.1140/epja/i2015-15173-1}.

\bibitem[{\citenamefont{Usang et~al.}(2019)\citenamefont{Usang, Ivanyuk,
  Ishizuka, and Chiba}}]{Usang2019}
\bibinfo{author}{\bibfnamefont{M.~D.} \bibnamefont{Usang}},
  \bibinfo{author}{\bibfnamefont{F.~A.} \bibnamefont{Ivanyuk}},
  \bibinfo{author}{\bibfnamefont{C.}~\bibnamefont{Ishizuka}}, \bibnamefont{and}
  \bibinfo{author}{\bibfnamefont{S.}~\bibnamefont{Chiba}},
  \bibinfo{journal}{Scientific Reports} \textbf{\bibinfo{volume}{9}},
  \bibinfo{pages}{1525} (\bibinfo{year}{2019}),
  \urlprefix\url{https://doi.org/10.1038/s41598-018-37993-7}.

\bibitem[{\citenamefont{Ishizuka et~al.}(2017)\citenamefont{Ishizuka, Usang,
  Ivanyuk, Maruhn, Nishio, and Chiba}}]{Ishizuka2017}
\bibinfo{author}{\bibfnamefont{C.}~\bibnamefont{Ishizuka}},
  \bibinfo{author}{\bibfnamefont{M.~D.} \bibnamefont{Usang}},
  \bibinfo{author}{\bibfnamefont{F.~A.} \bibnamefont{Ivanyuk}},
  \bibinfo{author}{\bibfnamefont{J.~A.} \bibnamefont{Maruhn}},
  \bibinfo{author}{\bibfnamefont{K.}~\bibnamefont{Nishio}}, \bibnamefont{and}
  \bibinfo{author}{\bibfnamefont{S.}~\bibnamefont{Chiba}},
  \bibinfo{journal}{Phys. Rev. C} \textbf{\bibinfo{volume}{96}},
  \bibinfo{pages}{064616} (\bibinfo{year}{2017}),
  \urlprefix\url{https://link.aps.org/doi/10.1103/PhysRevC.96.064616}.

\bibitem[{\citenamefont{Mumpower et~al.}(2020)\citenamefont{Mumpower, Jaffke,
  Verriere, and Randrup}}]{Mumpower2020}
\bibinfo{author}{\bibfnamefont{M.~R.} \bibnamefont{Mumpower}},
  \bibinfo{author}{\bibfnamefont{P.}~\bibnamefont{Jaffke}},
  \bibinfo{author}{\bibfnamefont{M.}~\bibnamefont{Verriere}}, \bibnamefont{and}
  \bibinfo{author}{\bibfnamefont{J.}~\bibnamefont{Randrup}},
  \bibinfo{journal}{Phys. Rev. C} \textbf{\bibinfo{volume}{101}},
  \bibinfo{pages}{054607} (\bibinfo{year}{2020}),
  \urlprefix\url{https://link.aps.org/doi/10.1103/PhysRevC.101.054607}.

\bibitem[{\citenamefont{Verriere et~al.}(2021)\citenamefont{Verriere, Schunck,
  and Regnier}}]{Verriere2021}
\bibinfo{author}{\bibfnamefont{M.}~\bibnamefont{Verriere}},
  \bibinfo{author}{\bibfnamefont{N.}~\bibnamefont{Schunck}}, \bibnamefont{and}
  \bibinfo{author}{\bibfnamefont{D.}~\bibnamefont{Regnier}},
  \bibinfo{journal}{Phys. Rev. C} \textbf{\bibinfo{volume}{103}},
  \bibinfo{pages}{054602} (\bibinfo{year}{2021}),
  \urlprefix\url{https://link.aps.org/doi/10.1103/PhysRevC.103.054602}.

\bibitem[{\citenamefont{Randrup and Vogt}(2014)}]{Randrup:2014}
\bibinfo{author}{\bibfnamefont{J.}~\bibnamefont{Randrup}} \bibnamefont{and}
  \bibinfo{author}{\bibfnamefont{R.}~\bibnamefont{Vogt}},
  \bibinfo{journal}{Phys. Rev. C} \textbf{\bibinfo{volume}{89}},
  \bibinfo{pages}{044601} (\bibinfo{year}{2014}),
  \urlprefix\url{https://link.aps.org/doi/10.1103/PhysRevC.89.044601}.

\bibitem[{\citenamefont{Marevi\ifmmode~\acute{c}\else \'{c}\fi{}
  et~al.}(2021)\citenamefont{Marevi\ifmmode~\acute{c}\else \'{c}\fi{}, Schunck,
  Randrup, and Vogt}}]{Marevic2021}
\bibinfo{author}{\bibfnamefont{P.}~\bibnamefont{Marevi\ifmmode~\acute{c}\else
  \'{c}\fi{}}}, \bibinfo{author}{\bibfnamefont{N.}~\bibnamefont{Schunck}},
  \bibinfo{author}{\bibfnamefont{J.}~\bibnamefont{Randrup}}, \bibnamefont{and}
  \bibinfo{author}{\bibfnamefont{R.}~\bibnamefont{Vogt}},
  \bibinfo{journal}{Phys. Rev. C} \textbf{\bibinfo{volume}{104}},
  \bibinfo{pages}{L021601} (\bibinfo{year}{2021}),
  \urlprefix\url{https://link.aps.org/doi/10.1103/PhysRevC.104.L021601}.

\bibitem[{\citenamefont{Bulgac et~al.}(2021)\citenamefont{Bulgac, Abdurrahman,
  Jin, Godbey, Schunck, and Stetcu}}]{Bulgac2021}
\bibinfo{author}{\bibfnamefont{A.}~\bibnamefont{Bulgac}},
  \bibinfo{author}{\bibfnamefont{I.}~\bibnamefont{Abdurrahman}},
  \bibinfo{author}{\bibfnamefont{S.}~\bibnamefont{Jin}},
  \bibinfo{author}{\bibfnamefont{K.}~\bibnamefont{Godbey}},
  \bibinfo{author}{\bibfnamefont{N.}~\bibnamefont{Schunck}}, \bibnamefont{and}
  \bibinfo{author}{\bibfnamefont{I.}~\bibnamefont{Stetcu}},
  \bibinfo{journal}{Phys. Rev. Lett.} \textbf{\bibinfo{volume}{126}},
  \bibinfo{pages}{142502} (\bibinfo{year}{2021}),
  \urlprefix\url{https://link.aps.org/doi/10.1103/PhysRevLett.126.142502}.

\bibitem[{\citenamefont{Randrup and Vogt}(2021)}]{Randrup:2021}
\bibinfo{author}{\bibfnamefont{J.}~\bibnamefont{Randrup}} \bibnamefont{and}
  \bibinfo{author}{\bibfnamefont{R.}~\bibnamefont{Vogt}},
  \bibinfo{journal}{Phys. Rev. Lett.} \textbf{\bibinfo{volume}{127}},
  \bibinfo{pages}{062502} (\bibinfo{year}{2021}),
  \urlprefix\url{https://link.aps.org/doi/10.1103/PhysRevLett.127.062502}.

\bibitem[{\citenamefont{Stetcu et~al.}(2021)\citenamefont{Stetcu, Lovell,
  Talou, Kawano, Marin, Pozzi, and Bulgac}}]{Stetcu2021}
\bibinfo{author}{\bibfnamefont{I.}~\bibnamefont{Stetcu}},
  \bibinfo{author}{\bibfnamefont{A.~E.} \bibnamefont{Lovell}},
  \bibinfo{author}{\bibfnamefont{P.}~\bibnamefont{Talou}},
  \bibinfo{author}{\bibfnamefont{T.}~\bibnamefont{Kawano}},
  \bibinfo{author}{\bibfnamefont{S.}~\bibnamefont{Marin}},
  \bibinfo{author}{\bibfnamefont{S.~A.} \bibnamefont{Pozzi}}, \bibnamefont{and}
  \bibinfo{author}{\bibfnamefont{A.}~\bibnamefont{Bulgac}},
  \bibinfo{journal}{Phys. Rev. Lett.} \textbf{\bibinfo{volume}{127}},
  \bibinfo{pages}{222502} (\bibinfo{year}{2021}),
  \urlprefix\url{https://link.aps.org/doi/10.1103/PhysRevLett.127.222502}.

\bibitem[{\citenamefont{Talou et~al.}(2021)\citenamefont{Talou, Stetcu, Jaffke,
  Rising, Lovell, and Kawano}}]{CGMF}
\bibinfo{author}{\bibfnamefont{P.}~\bibnamefont{Talou}},
  \bibinfo{author}{\bibfnamefont{I.}~\bibnamefont{Stetcu}},
  \bibinfo{author}{\bibfnamefont{P.}~\bibnamefont{Jaffke}},
  \bibinfo{author}{\bibfnamefont{M.}~\bibnamefont{Rising}},
  \bibinfo{author}{\bibfnamefont{A.}~\bibnamefont{Lovell}}, \bibnamefont{and}
  \bibinfo{author}{\bibfnamefont{T.}~\bibnamefont{Kawano}},
  \bibinfo{journal}{Comput. Phys. Commun.} \textbf{\bibinfo{volume}{269}},
  \bibinfo{pages}{108087} (\bibinfo{year}{2021}), ISSN
  \bibinfo{issn}{0010-4655},
  \urlprefix\url{https://www.sciencedirect.com/science/article/pii/S0010465521001995}.

\bibitem[{\citenamefont{Verbeke et~al.}(2015)\citenamefont{Verbeke, Randrup,
  and Vogt}}]{FREYA1}
\bibinfo{author}{\bibfnamefont{J.}~\bibnamefont{Verbeke}},
  \bibinfo{author}{\bibfnamefont{J.}~\bibnamefont{Randrup}}, \bibnamefont{and}
  \bibinfo{author}{\bibfnamefont{R.}~\bibnamefont{Vogt}},
  \bibinfo{journal}{Computer Physics Communications}
  \textbf{\bibinfo{volume}{191}}, \bibinfo{pages}{178 } (\bibinfo{year}{2015}),
  ISSN \bibinfo{issn}{0010-4655},
  \urlprefix\url{http://www.sciencedirect.com/science/article/pii/S0010465515000466}.

\bibitem[{\citenamefont{Verbeke et~al.}(2018)\citenamefont{Verbeke, Randrup,
  and Vogt}}]{FREYA2}
\bibinfo{author}{\bibfnamefont{J.}~\bibnamefont{Verbeke}},
  \bibinfo{author}{\bibfnamefont{J.}~\bibnamefont{Randrup}}, \bibnamefont{and}
  \bibinfo{author}{\bibfnamefont{R.}~\bibnamefont{Vogt}},
  \bibinfo{journal}{Computer Physics Communications}
  \textbf{\bibinfo{volume}{222}}, \bibinfo{pages}{263 } (\bibinfo{year}{2018}),
  ISSN \bibinfo{issn}{0010-4655},
  \urlprefix\url{http://www.sciencedirect.com/science/article/pii/S001046551730293X}.

\bibitem[{\citenamefont{Litaize et~al.}(2015)\citenamefont{Litaize, Serot, and
  Berge}}]{FIFRELIN}
\bibinfo{author}{\bibfnamefont{O.}~\bibnamefont{Litaize}},
  \bibinfo{author}{\bibfnamefont{O.}~\bibnamefont{Serot}}, \bibnamefont{and}
  \bibinfo{author}{\bibfnamefont{L.}~\bibnamefont{Berge}},
  \bibinfo{journal}{The European Physical Journal A}
  \textbf{\bibinfo{volume}{51}}, \bibinfo{pages}{177} (\bibinfo{year}{2015}),
  ISSN \bibinfo{issn}{1434-601X},
  \urlprefix\url{https://doi.org/10.1140/epja/i2015-15177-9}.

\bibitem[{\citenamefont{Okumura et~al.}(2018)\citenamefont{Okumura, Kawano,
  Talou, Jaffke, and Chiba}}]{Okumura2018}
\bibinfo{author}{\bibfnamefont{S.}~\bibnamefont{Okumura}},
  \bibinfo{author}{\bibfnamefont{T.}~\bibnamefont{Kawano}},
  \bibinfo{author}{\bibfnamefont{P.}~\bibnamefont{Talou}},
  \bibinfo{author}{\bibfnamefont{P.}~\bibnamefont{Jaffke}}, \bibnamefont{and}
  \bibinfo{author}{\bibfnamefont{S.}~\bibnamefont{Chiba}}, \bibinfo{journal}{J.
  Nucl. Sci. Tech.} \textbf{\bibinfo{volume}{55}}, \bibinfo{pages}{1009}
  (\bibinfo{year}{2018}).

\bibitem[{\citenamefont{Lovell et~al.}(2021)\citenamefont{Lovell, Kawano,
  Okumura, Stetcu, Mumpower, and Talou}}]{Lovell2021}
\bibinfo{author}{\bibfnamefont{A.~E.} \bibnamefont{Lovell}},
  \bibinfo{author}{\bibfnamefont{T.}~\bibnamefont{Kawano}},
  \bibinfo{author}{\bibfnamefont{S.}~\bibnamefont{Okumura}},
  \bibinfo{author}{\bibfnamefont{I.}~\bibnamefont{Stetcu}},
  \bibinfo{author}{\bibfnamefont{M.~R.} \bibnamefont{Mumpower}},
  \bibnamefont{and} \bibinfo{author}{\bibfnamefont{P.}~\bibnamefont{Talou}},
  \bibinfo{journal}{Phys. Rev. C} \textbf{\bibinfo{volume}{103}},
  \bibinfo{pages}{014615} (\bibinfo{year}{2021}),
  \urlprefix\url{https://link.aps.org/doi/10.1103/PhysRevC.103.014615}.

\bibitem[{\citenamefont{Schmidt et~al.}(2017)\citenamefont{Schmidt, Jurado, and
  Schmitt}}]{GEF}
\bibinfo{author}{\bibfnamefont{K.-H.} \bibnamefont{Schmidt}},
  \bibinfo{author}{\bibfnamefont{B.}~\bibnamefont{Jurado}}, \bibnamefont{and}
  \bibinfo{author}{\bibfnamefont{C.}~\bibnamefont{Schmitt}},
  \bibinfo{journal}{EPJ Web of Conferences} \textbf{\bibinfo{volume}{146}},
  \bibinfo{pages}{04001} (\bibinfo{year}{2017}).

\bibitem[{\citenamefont{Tudora and Hambsch}(2017)}]{Tudora2017}
\bibinfo{author}{\bibfnamefont{A.}~\bibnamefont{Tudora}} \bibnamefont{and}
  \bibinfo{author}{\bibfnamefont{F.~J.} \bibnamefont{Hambsch}},
  \bibinfo{journal}{The European Physical Journal A}
  \textbf{\bibinfo{volume}{53}}, \bibinfo{pages}{159} (\bibinfo{year}{2017}),
  \urlprefix\url{https://doi.org/10.1140/epja/i2017-12347-9}.

\bibitem[{\citenamefont{Brown et~al.}(2018)\citenamefont{Brown, Chadwick,
  Capote, Kahler, Trkov, Herman, Sonzogni, Danon, Carlson, Dunn
  et~al.}}]{ENDFB8}
\bibinfo{author}{\bibfnamefont{D.}~\bibnamefont{Brown}},
  \bibinfo{author}{\bibfnamefont{M.}~\bibnamefont{Chadwick}},
  \bibinfo{author}{\bibfnamefont{R.}~\bibnamefont{Capote}},
  \bibinfo{author}{\bibfnamefont{A.}~\bibnamefont{Kahler}},
  \bibinfo{author}{\bibfnamefont{A.}~\bibnamefont{Trkov}},
  \bibinfo{author}{\bibfnamefont{M.}~\bibnamefont{Herman}},
  \bibinfo{author}{\bibfnamefont{A.}~\bibnamefont{Sonzogni}},
  \bibinfo{author}{\bibfnamefont{Y.}~\bibnamefont{Danon}},
  \bibinfo{author}{\bibfnamefont{A.}~\bibnamefont{Carlson}},
  \bibinfo{author}{\bibfnamefont{M.}~\bibnamefont{Dunn}}, \bibnamefont{et~al.},
  \bibinfo{journal}{Nuclear Data Sheets} \textbf{\bibinfo{volume}{148}},
  \bibinfo{pages}{1 } (\bibinfo{year}{2018}), ISSN \bibinfo{issn}{0090-3752},
  \bibinfo{note}{special Issue on Nuclear Reaction Data},
  \urlprefix\url{http://www.sciencedirect.com/science/article/pii/S0090375218300206}.

\bibitem[{\citenamefont{Plompen et~al.}(2020)\citenamefont{Plompen, Cabellos,
  De~Saint~Jean, Fleming, Algora, Angelone, Archier, Bauge, Bersillon, Blokhin
  et~al.}}]{JEFF33}
\bibinfo{author}{\bibfnamefont{A.~J.~M.} \bibnamefont{Plompen}},
  \bibinfo{author}{\bibfnamefont{O.}~\bibnamefont{Cabellos}},
  \bibinfo{author}{\bibfnamefont{C.}~\bibnamefont{De~Saint~Jean}},
  \bibinfo{author}{\bibfnamefont{M.}~\bibnamefont{Fleming}},
  \bibinfo{author}{\bibfnamefont{A.}~\bibnamefont{Algora}},
  \bibinfo{author}{\bibfnamefont{M.}~\bibnamefont{Angelone}},
  \bibinfo{author}{\bibfnamefont{P.}~\bibnamefont{Archier}},
  \bibinfo{author}{\bibfnamefont{E.}~\bibnamefont{Bauge}},
  \bibinfo{author}{\bibfnamefont{O.}~\bibnamefont{Bersillon}},
  \bibinfo{author}{\bibfnamefont{A.}~\bibnamefont{Blokhin}},
  \bibnamefont{et~al.}, \bibinfo{journal}{The European Physical Journal A}
  \textbf{\bibinfo{volume}{56}}, \bibinfo{pages}{181} (\bibinfo{year}{2020}),
  \urlprefix\url{https://doi.org/10.1140/epja/s10050-020-00141-9}.

\bibitem[{\citenamefont{Iwamoto et~al.}(2023)\citenamefont{Iwamoto, Iwamoto,
  Kunieda, Minato, Nakayama, Abe, Tsubakihara, Okumura, Ishizuka, Yoshida
  et~al.}}]{JENDL5}
\bibinfo{author}{\bibfnamefont{O.}~\bibnamefont{Iwamoto}},
  \bibinfo{author}{\bibfnamefont{N.}~\bibnamefont{Iwamoto}},
  \bibinfo{author}{\bibfnamefont{S.}~\bibnamefont{Kunieda}},
  \bibinfo{author}{\bibfnamefont{F.}~\bibnamefont{Minato}},
  \bibinfo{author}{\bibfnamefont{S.}~\bibnamefont{Nakayama}},
  \bibinfo{author}{\bibfnamefont{Y.}~\bibnamefont{Abe}},
  \bibinfo{author}{\bibfnamefont{K.}~\bibnamefont{Tsubakihara}},
  \bibinfo{author}{\bibfnamefont{S.}~\bibnamefont{Okumura}},
  \bibinfo{author}{\bibfnamefont{C.}~\bibnamefont{Ishizuka}},
  \bibinfo{author}{\bibfnamefont{T.}~\bibnamefont{Yoshida}},
  \bibnamefont{et~al.}, \bibinfo{journal}{Journal of Nuclear Science and
  Technology} \textbf{\bibinfo{volume}{60}}, \bibinfo{pages}{1}
  (\bibinfo{year}{2023}),
  \urlprefix\url{https://doi.org/10.1080/00223131.2022.2141903}.

\bibitem[{\citenamefont{Hauser and Feshbach}(1952)}]{HauserFeshbach}
\bibinfo{author}{\bibfnamefont{W.}~\bibnamefont{Hauser}} \bibnamefont{and}
  \bibinfo{author}{\bibfnamefont{H.}~\bibnamefont{Feshbach}},
  \bibinfo{journal}{Phys. Rev.} \textbf{\bibinfo{volume}{87}},
  \bibinfo{pages}{366} (\bibinfo{year}{1952}),
  \urlprefix\url{https://link.aps.org/doi/10.1103/PhysRev.87.366}.

\bibitem[{\citenamefont{Novak et~al.}(2014)\citenamefont{Novak, Novak, Pratt,
  Vredevoogd, Coleman-Smith, and Wolpert}}]{Novak2014}
\bibinfo{author}{\bibfnamefont{J.}~\bibnamefont{Novak}},
  \bibinfo{author}{\bibfnamefont{K.}~\bibnamefont{Novak}},
  \bibinfo{author}{\bibfnamefont{S.}~\bibnamefont{Pratt}},
  \bibinfo{author}{\bibfnamefont{J.}~\bibnamefont{Vredevoogd}},
  \bibinfo{author}{\bibfnamefont{C.~E.} \bibnamefont{Coleman-Smith}},
  \bibnamefont{and} \bibinfo{author}{\bibfnamefont{R.~L.}
  \bibnamefont{Wolpert}}, \bibinfo{journal}{Phys. Rev. C}
  \textbf{\bibinfo{volume}{89}}, \bibinfo{pages}{034917}
  (\bibinfo{year}{2014}),
  \urlprefix\url{https://link.aps.org/doi/10.1103/PhysRevC.89.034917}.

\bibitem[{\citenamefont{Sangaline and Pratt}(2016)}]{Sangaline2016}
\bibinfo{author}{\bibfnamefont{E.}~\bibnamefont{Sangaline}} \bibnamefont{and}
  \bibinfo{author}{\bibfnamefont{S.}~\bibnamefont{Pratt}},
  \bibinfo{journal}{Phys. Rev. C} \textbf{\bibinfo{volume}{93}},
  \bibinfo{pages}{024908} (\bibinfo{year}{2016}),
  \urlprefix\url{https://link.aps.org/doi/10.1103/PhysRevC.93.024908}.

\bibitem[{\citenamefont{Melendez et~al.}(2019)\citenamefont{Melendez,
  Furnstahl, Phillips, Pratola, and Wesolowski}}]{Melendez2019}
\bibinfo{author}{\bibfnamefont{J.~A.} \bibnamefont{Melendez}},
  \bibinfo{author}{\bibfnamefont{R.~J.} \bibnamefont{Furnstahl}},
  \bibinfo{author}{\bibfnamefont{D.~R.} \bibnamefont{Phillips}},
  \bibinfo{author}{\bibfnamefont{M.~T.} \bibnamefont{Pratola}},
  \bibnamefont{and}
  \bibinfo{author}{\bibfnamefont{S.}~\bibnamefont{Wesolowski}},
  \bibinfo{journal}{Phys. Rev. C} \textbf{\bibinfo{volume}{100}},
  \bibinfo{pages}{044001} (\bibinfo{year}{2019}),
  \urlprefix\url{https://link.aps.org/doi/10.1103/PhysRevC.100.044001}.

\bibitem[{\citenamefont{S\"urer et~al.}(2022)\citenamefont{S\"urer, Nunes,
  Plumlee, and Wild}}]{Surer2022}
\bibinfo{author}{\bibfnamefont{O.}~\bibnamefont{S\"urer}},
  \bibinfo{author}{\bibfnamefont{F.~M.} \bibnamefont{Nunes}},
  \bibinfo{author}{\bibfnamefont{M.}~\bibnamefont{Plumlee}}, \bibnamefont{and}
  \bibinfo{author}{\bibfnamefont{S.~M.} \bibnamefont{Wild}},
  \bibinfo{journal}{Phys. Rev. C} \textbf{\bibinfo{volume}{106}},
  \bibinfo{pages}{024607} (\bibinfo{year}{2022}),
  \urlprefix\url{https://link.aps.org/doi/10.1103/PhysRevC.106.024607}.

\bibitem[{\citenamefont{Frame et~al.}(2018)\citenamefont{Frame, He, Ipsen, Lee,
  Lee, and Rrapaj}}]{Frame2018}
\bibinfo{author}{\bibfnamefont{D.}~\bibnamefont{Frame}},
  \bibinfo{author}{\bibfnamefont{R.}~\bibnamefont{He}},
  \bibinfo{author}{\bibfnamefont{I.}~\bibnamefont{Ipsen}},
  \bibinfo{author}{\bibfnamefont{D.}~\bibnamefont{Lee}},
  \bibinfo{author}{\bibfnamefont{D.}~\bibnamefont{Lee}}, \bibnamefont{and}
  \bibinfo{author}{\bibfnamefont{E.}~\bibnamefont{Rrapaj}},
  \bibinfo{journal}{Phys. Rev. Lett.} \textbf{\bibinfo{volume}{121}},
  \bibinfo{pages}{032501} (\bibinfo{year}{2018}),
  \urlprefix\url{https://link.aps.org/doi/10.1103/PhysRevLett.121.032501}.

\bibitem[{\citenamefont{Sarkar and Lee}(2021)}]{Sarkar2021}
\bibinfo{author}{\bibfnamefont{A.}~\bibnamefont{Sarkar}} \bibnamefont{and}
  \bibinfo{author}{\bibfnamefont{D.}~\bibnamefont{Lee}},
  \bibinfo{journal}{Phys. Rev. Lett.} \textbf{\bibinfo{volume}{126}},
  \bibinfo{pages}{032501} (\bibinfo{year}{2021}),
  \urlprefix\url{https://link.aps.org/doi/10.1103/PhysRevLett.126.032501}.

\bibitem[{\citenamefont{K\"onig et~al.}(2020)\citenamefont{K\"onig, Ekstr\"om,
  Hebeler, Lee, and Schwenk}}]{Koenig2020}
\bibinfo{author}{\bibfnamefont{S.}~\bibnamefont{K\"onig}},
  \bibinfo{author}{\bibfnamefont{A.}~\bibnamefont{Ekstr\"om}},
  \bibinfo{author}{\bibfnamefont{K.}~\bibnamefont{Hebeler}},
  \bibinfo{author}{\bibfnamefont{D.}~\bibnamefont{Lee}}, \bibnamefont{and}
  \bibinfo{author}{\bibfnamefont{A.}~\bibnamefont{Schwenk}},
  \bibinfo{journal}{Phys. Lett. B} \textbf{\bibinfo{volume}{810}},
  \bibinfo{pages}{135814} (\bibinfo{year}{2020}), ISSN
  \bibinfo{issn}{0370-2693},
  \urlprefix\url{https://www.sciencedirect.com/science/article/pii/S0370269320306171}.

\bibitem[{\citenamefont{Ekstr\"om and Hagen}(2019)}]{Ekstrom2019}
\bibinfo{author}{\bibfnamefont{A.}~\bibnamefont{Ekstr\"om}} \bibnamefont{and}
  \bibinfo{author}{\bibfnamefont{G.}~\bibnamefont{Hagen}},
  \bibinfo{journal}{Phys. Rev. Lett.} \textbf{\bibinfo{volume}{123}},
  \bibinfo{pages}{252501} (\bibinfo{year}{2019}),
  \urlprefix\url{https://link.aps.org/doi/10.1103/PhysRevLett.123.252501}.

\bibitem[{\citenamefont{Furnstahl et~al.}(2020)\citenamefont{Furnstahl, Garcia,
  Millican, and Zhang}}]{Furnstahl2020}
\bibinfo{author}{\bibfnamefont{R.}~\bibnamefont{Furnstahl}},
  \bibinfo{author}{\bibfnamefont{A.}~\bibnamefont{Garcia}},
  \bibinfo{author}{\bibfnamefont{P.}~\bibnamefont{Millican}}, \bibnamefont{and}
  \bibinfo{author}{\bibfnamefont{X.}~\bibnamefont{Zhang}},
  \bibinfo{journal}{Physics Letters B} \textbf{\bibinfo{volume}{809}},
  \bibinfo{pages}{135719} (\bibinfo{year}{2020}), ISSN
  \bibinfo{issn}{0370-2693},
  \urlprefix\url{https://www.sciencedirect.com/science/article/pii/S0370269320305220}.

\bibitem[{\citenamefont{Wesolowski et~al.}(2021)\citenamefont{Wesolowski,
  Svensson, Ekstr\"om, Forssén, Furnstahl, Melendez, and
  Phillips}}]{Wesolowski2021}
\bibinfo{author}{\bibfnamefont{S.}~\bibnamefont{Wesolowski}},
  \bibinfo{author}{\bibfnamefont{I.}~\bibnamefont{Svensson}},
  \bibinfo{author}{\bibfnamefont{A.}~\bibnamefont{Ekstr\"om}},
  \bibinfo{author}{\bibfnamefont{C.}~\bibnamefont{Forssén}},
  \bibinfo{author}{\bibfnamefont{R.~J.} \bibnamefont{Furnstahl}},
  \bibinfo{author}{\bibfnamefont{J.~A.} \bibnamefont{Melendez}},
  \bibnamefont{and} \bibinfo{author}{\bibfnamefont{D.~R.}
  \bibnamefont{Phillips}}, \emph{\bibinfo{title}{Fast \& rigorous constraints
  on chiral three-nucleon forces from few-body observables}}
  (\bibinfo{year}{2021}), \eprint{2104.04441}.

\bibitem[{\citenamefont{Utama et~al.}(2016{\natexlab{a}})\citenamefont{Utama,
  Chen, and Piekarewicz}}]{Utama2016}
\bibinfo{author}{\bibfnamefont{R.}~\bibnamefont{Utama}},
  \bibinfo{author}{\bibfnamefont{W.-C.} \bibnamefont{Chen}}, \bibnamefont{and}
  \bibinfo{author}{\bibfnamefont{J.}~\bibnamefont{Piekarewicz}},
  \bibinfo{journal}{Journal of Physics G: Nuclear and Particle Physics}
  \textbf{\bibinfo{volume}{43}}, \bibinfo{pages}{114002}
  (\bibinfo{year}{2016}{\natexlab{a}}),
  \urlprefix\url{http://stacks.iop.org/0954-3899/43/i=11/a=114002}.

\bibitem[{\citenamefont{Utama et~al.}(2016{\natexlab{b}})\citenamefont{Utama,
  Piekarewicz, and Prosper}}]{Utama2016a}
\bibinfo{author}{\bibfnamefont{R.}~\bibnamefont{Utama}},
  \bibinfo{author}{\bibfnamefont{J.}~\bibnamefont{Piekarewicz}},
  \bibnamefont{and} \bibinfo{author}{\bibfnamefont{H.~B.}
  \bibnamefont{Prosper}}, \bibinfo{journal}{Phys. Rev. C}
  \textbf{\bibinfo{volume}{93}}, \bibinfo{pages}{014311}
  (\bibinfo{year}{2016}{\natexlab{b}}),
  \urlprefix\url{https://link.aps.org/doi/10.1103/PhysRevC.93.014311}.

\bibitem[{\citenamefont{Neufcourt et~al.}(2018)\citenamefont{Neufcourt, Cao,
  Nazarewicz, and Viens}}]{Neufcourt2018}
\bibinfo{author}{\bibfnamefont{L.}~\bibnamefont{Neufcourt}},
  \bibinfo{author}{\bibfnamefont{Y.}~\bibnamefont{Cao}},
  \bibinfo{author}{\bibfnamefont{W.}~\bibnamefont{Nazarewicz}},
  \bibnamefont{and} \bibinfo{author}{\bibfnamefont{F.}~\bibnamefont{Viens}},
  \bibinfo{journal}{Phys. Rev. C} \textbf{\bibinfo{volume}{98}},
  \bibinfo{pages}{034318} (\bibinfo{year}{2018}),
  \urlprefix\url{https://link.aps.org/doi/10.1103/PhysRevC.98.034318}.

\bibitem[{\citenamefont{Wang et~al.}(2019)\citenamefont{Wang, Pei, Liu, and
  Qiang}}]{Wang2019}
\bibinfo{author}{\bibfnamefont{Z.-A.} \bibnamefont{Wang}},
  \bibinfo{author}{\bibfnamefont{J.}~\bibnamefont{Pei}},
  \bibinfo{author}{\bibfnamefont{Y.}~\bibnamefont{Liu}}, \bibnamefont{and}
  \bibinfo{author}{\bibfnamefont{Y.}~\bibnamefont{Qiang}},
  \bibinfo{journal}{Phys. Rev. Lett.} \textbf{\bibinfo{volume}{123}},
  \bibinfo{pages}{122501} (\bibinfo{year}{2019}),
  \urlprefix\url{https://link.aps.org/doi/10.1103/PhysRevLett.123.122501}.

\bibitem[{\citenamefont{Neufcourt et~al.}(2020)\citenamefont{Neufcourt, Cao,
  Giuliani, Nazarewicz, Olsen, and Tarasov}}]{Neufcourt2020a}
\bibinfo{author}{\bibfnamefont{L.}~\bibnamefont{Neufcourt}},
  \bibinfo{author}{\bibfnamefont{Y.}~\bibnamefont{Cao}},
  \bibinfo{author}{\bibfnamefont{S.~A.} \bibnamefont{Giuliani}},
  \bibinfo{author}{\bibfnamefont{W.}~\bibnamefont{Nazarewicz}},
  \bibinfo{author}{\bibfnamefont{E.}~\bibnamefont{Olsen}}, \bibnamefont{and}
  \bibinfo{author}{\bibfnamefont{O.~B.} \bibnamefont{Tarasov}},
  \bibinfo{journal}{Phys. Rev. C} \textbf{\bibinfo{volume}{101}},
  \bibinfo{pages}{044307} (\bibinfo{year}{2020}),
  \urlprefix\url{https://link.aps.org/doi/10.1103/PhysRevC.101.044307}.

\bibitem[{\citenamefont{Drischler et~al.}(2021)\citenamefont{Drischler,
  Quinonez, Giuliani, Lovell, and Nunes}}]{Drischler2021}
\bibinfo{author}{\bibfnamefont{C.}~\bibnamefont{Drischler}},
  \bibinfo{author}{\bibfnamefont{M.}~\bibnamefont{Quinonez}},
  \bibinfo{author}{\bibfnamefont{P.}~\bibnamefont{Giuliani}},
  \bibinfo{author}{\bibfnamefont{A.}~\bibnamefont{Lovell}}, \bibnamefont{and}
  \bibinfo{author}{\bibfnamefont{F.}~\bibnamefont{Nunes}},
  \bibinfo{journal}{Physics Letters B} \textbf{\bibinfo{volume}{823}},
  \bibinfo{pages}{136777} (\bibinfo{year}{2021}), ISSN
  \bibinfo{issn}{0370-2693},
  \urlprefix\url{https://www.sciencedirect.com/science/article/pii/S0370269321007176}.

\bibitem[{\citenamefont{Odell et~al.}(2024)\citenamefont{Odell, Giuliani,
  Beyer, Catacora-Rios, Chan, Bonilla, Furnstahl, Godbey, and
  Nunes}}]{Odell2024}
\bibinfo{author}{\bibfnamefont{D.}~\bibnamefont{Odell}},
  \bibinfo{author}{\bibfnamefont{P.}~\bibnamefont{Giuliani}},
  \bibinfo{author}{\bibfnamefont{K.}~\bibnamefont{Beyer}},
  \bibinfo{author}{\bibfnamefont{M.}~\bibnamefont{Catacora-Rios}},
  \bibinfo{author}{\bibfnamefont{M.~Y.-H.} \bibnamefont{Chan}},
  \bibinfo{author}{\bibfnamefont{E.}~\bibnamefont{Bonilla}},
  \bibinfo{author}{\bibfnamefont{R.~J.} \bibnamefont{Furnstahl}},
  \bibinfo{author}{\bibfnamefont{K.}~\bibnamefont{Godbey}}, \bibnamefont{and}
  \bibinfo{author}{\bibfnamefont{F.~M.} \bibnamefont{Nunes}},
  \bibinfo{journal}{Phys. Rev. C} \textbf{\bibinfo{volume}{109}},
  \bibinfo{pages}{044612} (\bibinfo{year}{2024}),
  \urlprefix\url{https://link.aps.org/doi/10.1103/PhysRevC.109.044612}.

\bibitem[{\citenamefont{Bishop}(1994)}]{MDNs}
\bibinfo{author}{\bibfnamefont{C.~M.} \bibnamefont{Bishop}},
  \bibinfo{type}{Tech. Rep.}, \bibinfo{institution}{Aston University,
  Department of Computer Science and Applied Mathematics}
  (\bibinfo{year}{1994}).

\bibitem[{\citenamefont{Lovell et~al.}(2020{\natexlab{a}})\citenamefont{Lovell,
  Mohan, and Talou}}]{Lovell2020}
\bibinfo{author}{\bibfnamefont{A.~E.} \bibnamefont{Lovell}},
  \bibinfo{author}{\bibfnamefont{A.~T.} \bibnamefont{Mohan}}, \bibnamefont{and}
  \bibinfo{author}{\bibfnamefont{P.}~\bibnamefont{Talou}},
  \bibinfo{journal}{Journal of Physics G: Nuclear and Particle Physics}
  \textbf{\bibinfo{volume}{47}}, \bibinfo{pages}{114001}
  (\bibinfo{year}{2020}{\natexlab{a}}),
  \urlprefix\url{https://doi.org/10.1088\%2F1361-6471\%2Fab9f58}.

\bibitem[{\citenamefont{Lovell et~al.}(2022)\citenamefont{Lovell, Mohan,
  Sprouse, and Mumpower}}]{Lovell2022}
\bibinfo{author}{\bibfnamefont{A.~E.} \bibnamefont{Lovell}},
  \bibinfo{author}{\bibfnamefont{A.~T.} \bibnamefont{Mohan}},
  \bibinfo{author}{\bibfnamefont{T.~M.} \bibnamefont{Sprouse}},
  \bibnamefont{and} \bibinfo{author}{\bibfnamefont{M.~R.}
  \bibnamefont{Mumpower}}, \bibinfo{journal}{Physical Review. C}
  \textbf{\bibinfo{volume}{106}} (\bibinfo{year}{2022}), ISSN
  \bibinfo{issn}{2469-9985},
  \urlprefix\url{https://www.osti.gov/biblio/1878054}.

\bibitem[{\citenamefont{Paszke et~al.}(2019)\citenamefont{Paszke, Gross, Massa,
  Lerer, Bradbury, Chanan, Killeen, Lin, Gimelshein, Antiga et~al.}}]{pytorch}
\bibinfo{author}{\bibfnamefont{A.}~\bibnamefont{Paszke}},
  \bibinfo{author}{\bibfnamefont{S.}~\bibnamefont{Gross}},
  \bibinfo{author}{\bibfnamefont{F.}~\bibnamefont{Massa}},
  \bibinfo{author}{\bibfnamefont{A.}~\bibnamefont{Lerer}},
  \bibinfo{author}{\bibfnamefont{J.}~\bibnamefont{Bradbury}},
  \bibinfo{author}{\bibfnamefont{G.}~\bibnamefont{Chanan}},
  \bibinfo{author}{\bibfnamefont{T.}~\bibnamefont{Killeen}},
  \bibinfo{author}{\bibfnamefont{Z.}~\bibnamefont{Lin}},
  \bibinfo{author}{\bibfnamefont{N.}~\bibnamefont{Gimelshein}},
  \bibinfo{author}{\bibfnamefont{L.}~\bibnamefont{Antiga}},
  \bibnamefont{et~al.}, in \emph{\bibinfo{booktitle}{Advances in Neural
  Information Processing Systems 32}}, edited by
  \bibinfo{editor}{\bibfnamefont{H.}~\bibnamefont{Wallach}},
  \bibinfo{editor}{\bibfnamefont{H.}~\bibnamefont{Larochelle}},
  \bibinfo{editor}{\bibfnamefont{A.}~\bibnamefont{Beygelzimer}},
  \bibinfo{editor}{\bibfnamefont{F.}~\bibnamefont{d\textquotesingle
  Alch\'e-Buc}}, \bibinfo{editor}{\bibfnamefont{E.}~\bibnamefont{Fox}},
  \bibnamefont{and} \bibinfo{editor}{\bibfnamefont{R.}~\bibnamefont{Garnett}}
  (\bibinfo{publisher}{Curran Associates, Inc.}, \bibinfo{year}{2019}), pp.
  \bibinfo{pages}{8024--8035},
  \urlprefix\url{http://papers.neurips.cc/paper/9015-pytorch-an-imperative-style-high-performance-deep-learning-library.pdf}.

\bibitem[{\citenamefont{Talou}(2023)}]{Talou2023}
\bibinfo{author}{\bibfnamefont{P.}~\bibnamefont{Talou}},
  \bibinfo{howpublished}{private communication} (\bibinfo{year}{2023}).

\bibitem[{\citenamefont{Stetcu et~al.}(2014)\citenamefont{Stetcu, Talou,
  Kawano, and Jandel}}]{Stetcu2014}
\bibinfo{author}{\bibfnamefont{I.}~\bibnamefont{Stetcu}},
  \bibinfo{author}{\bibfnamefont{P.}~\bibnamefont{Talou}},
  \bibinfo{author}{\bibfnamefont{T.}~\bibnamefont{Kawano}}, \bibnamefont{and}
  \bibinfo{author}{\bibfnamefont{M.}~\bibnamefont{Jandel}},
  \bibinfo{journal}{Phys. Rev. C} \textbf{\bibinfo{volume}{90}},
  \bibinfo{pages}{024617} (\bibinfo{year}{2014}),
  \urlprefix\url{https://link.aps.org/doi/10.1103/PhysRevC.90.024617}.

\bibitem[{\citenamefont{Lovell et~al.}(2020{\natexlab{b}})\citenamefont{Lovell,
  Talou, Stetcu, and Kelly}}]{Lovell2020a}
\bibinfo{author}{\bibfnamefont{A.~E.} \bibnamefont{Lovell}},
  \bibinfo{author}{\bibfnamefont{P.}~\bibnamefont{Talou}},
  \bibinfo{author}{\bibfnamefont{I.}~\bibnamefont{Stetcu}}, \bibnamefont{and}
  \bibinfo{author}{\bibfnamefont{K.~J.} \bibnamefont{Kelly}},
  \bibinfo{journal}{Phys. Rev. C} \textbf{\bibinfo{volume}{102}},
  \bibinfo{pages}{024621} (\bibinfo{year}{2020}{\natexlab{b}}),
  \urlprefix\url{https://link.aps.org/doi/10.1103/PhysRevC.102.024621}.

\bibitem[{\citenamefont{Kulesza et~al.}(2022)\citenamefont{Kulesza, Adams,
  Armstrong, Bolding, Brown, Bull, Burke, Clark, Forster, Giron et~al.}}]{MCNP}
\bibinfo{author}{\bibfnamefont{J.~A.} \bibnamefont{Kulesza}},
  \bibinfo{author}{\bibfnamefont{T.~R.} \bibnamefont{Adams}},
  \bibinfo{author}{\bibfnamefont{J.~C.} \bibnamefont{Armstrong}},
  \bibinfo{author}{\bibfnamefont{S.~R.} \bibnamefont{Bolding}},
  \bibinfo{author}{\bibfnamefont{F.~B.} \bibnamefont{Brown}},
  \bibinfo{author}{\bibfnamefont{J.~S.} \bibnamefont{Bull}},
  \bibinfo{author}{\bibfnamefont{T.~P.} \bibnamefont{Burke}},
  \bibinfo{author}{\bibfnamefont{A.~R.} \bibnamefont{Clark}},
  \bibinfo{author}{\bibfnamefont{R.~A.} \bibnamefont{Forster},
  \bibfnamefont{III}}, \bibinfo{author}{\bibfnamefont{J.~F.}
  \bibnamefont{Giron}}, \bibnamefont{et~al.}, \bibinfo{type}{Tech. Rep.}
  \bibinfo{number}{LA-UR-22-30006, Rev.~1}, \bibinfo{institution}{Los Alamos
  National Laboratory}, \bibinfo{address}{Los Alamos, NM, USA}
  (\bibinfo{year}{2022}), \urlprefix\url{https://www.osti.gov/biblio/1889957}.

\end{thebibliography}
\end{document}